\documentstyle[psfig,referee]{mn}
\newcommand{\bm}[1]{\mbox{\boldmath $#1$}}
\newcommand{\bdot}{\mbox{${\bf  \: \cdot \:}$}}
\newcommand{\smaller}{\scriptscriptstyle}
\newcommand{\btimes}{\mbox{${\bf \: \times \:}$}}

\newcommand{\hatn}{\hat{\bm{n}}}
\newcommand{\hatb}{\hat{\bm{b}}}
\newcommand{\bbet}{\bm{\beta}}
\newcommand{\bpar}{\beta_{\smaller \parallel}}
\newcommand{\bperp}{\bbet_{\smaller \perp}}
\newcommand{\evec}[1]{\mbox{$\bm{e}_{\rm #1}$}}

\newcommand{\tbu}{\widetilde{\beta}_{\rm u}}
\newcommand{\tbd}{\widetilde{\beta}_{\rm d}}

\newcommand{\etal}{{\rm {et al}.\ }}

\newcommand{\beq}{\begin{equation}}
\newcommand{\eeq}{\end{equation}}
\newcommand{\half}{\mbox{$\frac{1}{2}$}}
\newcommand{\third}{\mbox{$\frac{1}{3}$}}
\newcommand{\Gs}{\Gamma_{\rm s}}

\newcommand{\bs}{\beta_{\rm s}}
\newcommand{\bu}{\beta_{\rm u}}
\newcommand{\bd}{\beta_{\rm d}}
\newcommand{\brel}{\beta_{\rm rel}}
\newcommand{\Grel}{\Gamma_{\rm rel}}
\newcommand{\bsd}{\overline{\beta}_{\rm s}}
\newcommand{\barE}{\overline{E}}

\newcommand{\barm}{\overline{\mu}}

\newcommand{\Ei}{E_{\rm i}}
\newcommand{\Ef}{E_{\rm f}}

\newcommand{\bEi}{\barE_{\rm i}}
\newcommand{\bEf}{\barE_{\rm f}}

\newcommand{\barmu}{\overline{\mu}}

\newcommand{\mud}{\mu_{{\smaller\rightarrow}\rm d}}
\newcommand{\muu}{\mu_{{\smaller\rightarrow}\rm u}}
\newcommand{\bmud}{\bar{\mu}_{{\smaller\rightarrow}\rm d}}
\newcommand{\bmuu}{\bar{\mu}_{{\smaller\rightarrow}\rm u}}
\newcommand{\thd}{\theta_{{\smaller\rightarrow}\rm d}}
\newcommand{\thu}{\theta_{{\smaller\rightarrow}\rm u}}
\newcommand{\bthd}{\overline{\theta}_{{\smaller\rightarrow}\rm d}}
\newcommand{\bthu}{\overline{\theta}_{{\smaller\rightarrow}\rm u}}

\newcommand{\Bprp}{B_{\smaller\perp}}
\newcommand{\Bpll}{B_{\smaller\parallel}}
\newcommand{\lc}{\ell_{\rm c}}
\newcommand{\rg}{r_{\rm g}}

\newcommand{\Rs}{R_{\rm s}}
\newcommand{\Vs}{V_{\rm s}}

\newcommand{\Emax}{E_{\rm max}}

\newcommand{\Pret}{{\cal P}_{\rm ret}}
\newcommand{\Pesc}{{\cal P}_{\rm esc}}
\newcommand{\grad}{\bm{\nabla}}
\newcommand{\gesim}{\,\raisebox{-0.4ex}{$\stackrel{>}{\scriptstyle\sim}$}\,}
\newcommand{\lesim}{\,\raisebox{-0.4ex}{$\stackrel{<}{\scriptstyle\sim}$}\,}
\newcommand{\nskip}{\vskip \baselineskip \noindent}
\begin{document}
\title[Particle acceleration]
      {Particle acceleration by ultra-relativistic shocks: theory and simulations}

\author[Achterberg et al.]
       {Abraham Achterberg$^{1,2}$, Yves A. Gallant$^1$ 
       John. G. Kirk$^3$ and Axel W. Guthmann$^3$\\
        $^1$Astronomical Institute, Utrecht University,
            P.O.\ Box 80\,000, 3508 TA Utrecht, The Netherlands \\
        $^2$Center for High Energy Astrophysics, Kruislaan 403, 
           1098 SJ Amsterdam, The Netherlands\\
        $^3$ Max-Planck Institut f\"ur Kernphysik, 
        Postfach 10 39 80, 69029 Heidelberg, Germany}

\maketitle

\begin{abstract}
We consider the acceleration of charged particles near ultra-relativistic
shocks, with Lorentz factor $\Gs \gg 1$.  We present simulations of the 
acceleration process and compare these with results from semi-analytical 
calculations. We show that the spectrum that results from acceleration 
near ultra-relativistic shocks is a  power law, $N(E) \propto E^{-s}$, with a
nearly universal value $s \approx 2.2 - 2.3$ for the slope of this power law.

We confirm that the ultra-relativistic 
equivalent of Fermi acceleration at a shock differs from its non-relativistic 
counterpart by the occurence of large anisotropies in the distribution of the 
accelerated particles near the shock. In the rest frame of the upstream fluid, 
particles can only outrun the shock when their direction of motion lies within 
a small {\em loss cone} of opening angle $\theta_{\rm c} \approx \Gs^{-1}$ 
around the shock normal. 

We also show that all physically plausible deflection or scattering 
mechanisms can change the upstream flight direction of relativistic 
particles originating from downstream by only a small amount: 
$\Delta \theta \sim \Gs^{-1}$. This limits
the energy change per shock crossing cycle to $\Delta E \sim E$, except
for the first cycle where particles originate upstream. In that case
the upstream energy is boosted by a factor $\sim \Gs^{2}$ for those particles
that are scattered back across the shock into the upstream
region.

\end{abstract}

\begin{keywords}
 cosmic rays -- acceleration of particles -- shock waves --
 gamma-ray bursts.
\end{keywords}

\section{Introduction}

Diffusive shock acceleration (DSA), independently proposed
by various authors in the context of the acceleration of Galactic cosmic rays
(Krymskii, 1977; Axford, Leer \& Skadron, 1978;  Bell, 1978 and 
Blandford \& Ostriker, 1978), has become a paradigm for the production of
energetic particles near strong shocks in a magnetized plasma. 
It is believed to be the mechanism responsible for the production
of Galactic cosmic rays with an energy $E \le 10^{15}$ eV  
near the Sedov-Taylor blast waves associated with supernova remnants, 
below the so-called `knee' in the cosmic ray spectrum. 

Diffusive shock acceleration relies on repeated scattering of charged
particles by magnetic irregularities (Alfv\'en waves) to confine the particles
for some time near the shocks. This limits the mean free path of these particles
to values much less than the one derived for cosmic rays in the general interstellar
medium, $\lambda_{\rm ism} \sim 0.2$ pc for protons at $E \approx 5$ GeV, but
still significantly larger than the thickness of the shock.
This may explain the good correspondence between the radio maps of some supernova
remnants, which show the distribution of the GeV electrons and the magnetic field
responsible for the synchrotron emissivity, and the X-ray observations which show
the distribution of the hot $T \sim 10^{8}$ K shocked interstellar gas
(Achterberg, Blandford \& Reynolds, 1994; Aschenbach \& Leahy, 1999).
Direct observational evidence for particle acceleration
seems to be provided by the observation of 
X-ray synchrotron emission from electrons with $E \sim 100$ TeV in SN1006
(Koyama et al., 1995).

One of the main strengths of diffusive shock acceleration is that it predicts a
power-law spectrum which is not far from that required for a theory of
Galactic cosmic rays. The observed spectrum of these particles, 
$N_{\rm obs}(E) \: {\rm d}E \propto E^{-2.7} \; {\rm d}E$, is thought to arise from the
effects of energy-dependent propagation on a source spectrum in the range 
$N_{\rm s}(E) \: {\rm d}E \propto E^{-s} \; {\rm d}E$ with power-law index $s$
in the range $2.1 \lesim s \lesim 2.4$.
A similar spectrum is inferred for the GeV electrons responsible for 
the synchrotron emission in such non-thermal radio sources as supernova remnants 
and the lobes of radio galaxies, where the slope of the power law 
lies in the  range $s \approx 2 -2.7$.

The formation of a power-law spectrum is the result of the competition between 
the energy gain per shock crossing cycle, say from upstream to downstream and back,
and the chance of escape from the shock per crossing cycle, $\Pesc$.  
Shock acceleration therefore
is  a realization of the statistical acceleration process proposed by Fermi 
(1949).  The slope of the resulting power-law distribution is given by
\begin{equation}
\label{Fermislope}
	s = 1 + \frac{\ln(1/{\cal P}_{\rm ret})}{\ln \left< \Ef/\Ei \right>}
\end{equation}
Here $\Pret = 1 - \Pesc$ is the mean probability per cycle that a particle 
returns to the shock and re-crosses into the upstream medium,
$\Ef/\Ei$ is the ratio of final and initial energy in a cycle,  
and the angular brackets indicate the average value. 

For non-relativistic shocks one can use simple kinematics,
assuming elastic scattering of the accelerated particles in the upstream and 
downstream fluid rest frames, which leads to near-isotropic particle distributions. 
The calculation of the mean energy gain and return probability is straightforward,
and the resulting  slope $s$ depends only on the compression ratio 
$r = \rho_{\rm d}/\rho_{\rm u}$ at the shock (e.g. Bell, 1978):
\begin{equation}
\label{comprslope}
	s = \frac{r + 2}{r - 1} \; .
\end{equation}
Here $\rho_{\rm u}$ ($\rho_{\rm d}$) is the fluid density just upstream 
(downstream) of the shock.
For a strong non-relativistic shock in an ideal mon-atomic 
gas one has $r = 4$ and $s = 2$. The fact
that the observed spectra are often steeper is usually explained as the
effect of energy-dependent escape of accelerated particles from the source
(or the galaxy as a whole), or the modification of the shock by the back-reaction
of the accelerated particles: the gradient in the pressure of the accelerated 
particles slows down and heats the incoming fluid, 
decreasing the shock compression ratio. 

Reviews of the theory of diffusive shock acceleration
at non-relativistic shocks can be found in Drury (1983), Blandford \& Eichler (1987),
Jones \& Ellison (1991), Achterberg (1993) and Kirk (1994).

In this paper, we consider the process of Fermi-type shock acceleration in the
limit $\Gs \gg 1$. 
We briefly review the importance of such shocks in astrophysics in Section 2.
The basic kinematic constraints are discussed in Section 3.
There we will show that the energy gain per shock crossing is of order unity,
$\left< \Ef/\Ei \right> \approx 2$, except for the first shock crossing
(Gallant \& Achterberg, 1999). Possible scattering mechanisms are discussed
in Section 4, and the maximum energy that can be achieved with and without
losses is discussed in Section 5.
We present numerical simulations of the
acceleration process in Section 6, and compare these with semi-analytical results
of Kirk \etal (2000). Conclusions are presented in Section 7.

\section{Importance of relativistic shocks}

In this paper we consider particle acceleration at ultra-relativistic
shocks with shock velocity $V_{\rm s} = \bs \: c$ and bulk Lorentz factor 
$\Gs = (1 - \bs^{2})^{-1/2} \gg 1$. The importance of such shocks for 
particle acceleration to energies in the EeV range ($1$ EeV $= 10^{18}$ eV)
was pointed out by Hillas (1984). 
Quite general arguments, confirmed in our calculations below, show that
the maximum particle energy for a particle with charge $q = Ze$
that can be produced in a bulk magnetised
flow on a scale $\Rs$ with velocity $\bs c$ and magnetic field $B$ is
\[
	\Emax = ZeB  \: \Gs \bs \Rs \; .
\]
This value for $\Emax$ is a factor $\Gs$ larger than the one obtained from
the requirement that the particle gyration radius
is roughly the size of the system, $r_{\rm g}(E) \approx E/ZeB \sim \Rs$,
which is the criterion used by Hillas (1984). This difference is due to the
fact that, upstream, the particle typically completes only a fraction
of a gyro-orbit in a regular field, corresponding to an angle 
$\Delta \theta \sim 1/\Gs$, as is explained in Section 4.
This means that relativistic shocks seem to be the natural site of particle
acceleration to extreme energies.

First model calculations of acceleration at relativistic shocks
in the context of diffusive shock acceleration were 
done by Peacock (1981). He noted that 
for relativistic shock speeds the effect of relativistic beaming on the
angular distribution of accelerated particles near the shock becomes important.
This distribution determines the energy gain and the escape probability 
per crossing cycle, the quantities that determine the slope 
of the distribution of the accelerated particles. 

This conclusion has subsequently been confirmed by semi-analytical calculations 
for shocks with $\Gs < 10$ by Kirk \& Schneider (1987a/b) and
Schneider \& Kirk (1989). These authors use an eigenfunction method to 
describe the particle distributions on both sides of the shock, and determine
the slope of the spectrum from the continuity of the microscopic distribution
function across the shock. Heavens and Drury (1988) use a similar method, and their
results show a spectral index $s \approx 2.1 - 2.3$ for $\bs \approx 0.98$, which 
only depends weakly on the precise choice of the scattering operator employed in
their calculations. 
The eigenfunction method was recently extended to the case of ultra-relativistic 
shocks with $\Gs \gg 1$ by Kirk et al. (2000). These results will serve as a independent
check of the numerical results presented below. 

A renewed interest in ultra-relativistic shocks as sites of efficient particle 
acceleration has been sparked by key observations in two areas:
\begin{itemize}
\item	The observation of Ultra-High Energy Cosmic Rays (UHECRs)
	in the energy range $E \sim 3-300$ EeV, e.g. Bird \etal (1994); 
	Yoshida \etal (1995) and Takeda \etal (1998). 
	Assuming UHECRs are protons or light nuclei,  these particles are not 
	confined by the magnetic field of our Galaxy, 
	and should be of extragalactic origin given the fact that arrival
	directions do not cluster around the Galactic plane. Their
	high energy implies extraordinary circumstances in the production
	sites.
	 
	UHECRs are observed well above the 
	Greisen-Zatsepin-Kuz'min cut-off energy 
	$E_{\rm GZK} \approx 30$ EeV, with some 10 events above $10^{20}$ eV.
	$E_{\rm GZK}$ corresponds to the energy where losses 
	due to photo-pion production on the
	photons of the Cosmic Microwave Background become severe
	(Greisen, 1966; Zatsepin \& Kuz'min, 1966). This    
	suggests that the sources of UHECRs must be within a distance of 
	$\sim 50$ Mpc from the Galaxy, unless UHECRs are in fact created
	at much higher energies in a `top-down' scenario without acceleration, 
	for instance as the decay products of 
	some exotic particle (e.g. Farrar \& Biermann, 1998)
	or some quantum-mechanical topological 
	defect such as superconducting strings, e.g. Sigl \etal (1994). 
	A review of the relevant observations and
	theoretical considerations relating to UHECRs can be found in
	Bhattacharjee \& Sigl (2000). 

\vskip \baselineskip
  
\item	The observation of the X-ray, optical and radio afterglows of 
	Gamma Ray Bursts (GRBs), e.g. van Paradijs \etal (1997); 
	Djorgovski \etal (1997);
	Metzger \etal (1997); Frail \etal (1997), which proved that GRBs 
	originate at cosmological distances, as originally proposed by 
	Paczy\'nski (1986) and by Goodman (1986).
	The afterglow emission is believed to be Lorentz-boosted synchrotron 
	emission from relativistic electrons (e.g. Sari, Piran \& Narayan, 1998).
	These electrons must be accelerated at the external blast wave
	associated with the expanding GRB fireball, which is believed
	to have a bulk Lorentz factor 
	$\Gs \approx 100-1000$  (Cavallo \& Rees, 1978; Shemi \& Piran, 1990; 
 	M\'esz\'aros \& Rees, 1992a/b/c, 1993; Piran, Shemi \& Narayan, 1993).  
 	The short-duration (0.01-100 seconds)
	gamma-ray flash itself is believed to originate
	at internal shocks with Lorentz factor $\Gs \approx 2 - 10$, 
	e.g. Rees \& M\'esz\'aros, 1994; Sari \& Piran, 1995; 
	Sari \& Piran, 1997.
	These internal shocks are presumably 
	generated by inhomogeneities in the fireball or
	by fluctuations in the power output of the 
	`central engine' responsible for the GRB phenomenon.
	A recent review of  the GRB phenomenon and its possible implications
	can be found in Piran (1999). 
\end{itemize}	
A connection between these two phenomena has been suggested by Waxman (1995), 
who noted that observed UHECR flux above $100$ EeV and the mean gamma-ray flux due 
to GRBs are similar,  by Vietri (1995)  who suggested that the production of
UHECRs at the fireball blast wave could be extremely efficient, 
with an energy gain corresponding to 
$\left< \Ef/\Ei \right> \approx \Gs^{2} \gg 1$
per crossing cycle, and by Milgrom \& Usov (1995).	

Other sites where particles can be accelerated to EeV energies are the strong 
shocks in the hot spots of powerful (Fanaroff-Riley Class II) radio galaxies
(Rachen \& Biermann, 1993; Norman, Melrose \& Achterberg, 1995) and in the
internal shocks in the jets of Blazars.
These shocks are expected to be mildly relativistic ($\Gs \approx 2-10$).
Production scenarios involving non-relativistic but very large shocks associated 
with Large Scale Structure have also been proposed as a source of UHECRs 
(Kang, Ryu \& Jones, 1996).

\section{ENERGY GAIN AT RELATIVISTIC SHOCKS}

We consider a charged particle interacting with a shock with Lorentz factor
$\Gs \gg 1$ relative to the upstream medium. It is assumed that the upstream 
medium contains magnetic fields and possibly magnetic fluctuations which 
deflect or scatter charged particles in the upstream flow. 
In the case of ultra-relativistic 
shocks it is essential that strong magnetic fluctuations are present downstream 
so that particles can be magnetically scattered onto trajectories that allow 
them to return to the shock. This point will be considered in the discussion of 
Section 5.1. Under these circumstances a form of shock acceleration will operate 
with the same general principles as in the non-relativistic case. 

As we will show, particle acceleration near relativistic shocks is not 
{\em diffusive} shock acceleration because the propagation of 
accelerated particles near the shock, and in particular ahead of the shock, 
cannot be described as {\em spatial} diffusion. The anisotropies in
the angular distribution of the accelerated particles is large, and the diffusion
approximation for spatial transport does not apply. 

We consider a simple one-dimensional flow along the $z$-axis. We use units
where $c = 1$. 
We will have occasion to use three different frames of reference: the
{\em upstream rest frame} (URF), the {\em downstream rest frame} (DRF)
and the {\em shock rest frame} (SRF). 
Table 1 gives the notation used for various quantities in these
frames, and the relations between these quantities as follow from
the Lorentz transformations between these three frames of reference.
A bar ($\overline{\cdots}$) is used to denote quantities in the DRF, 
and a tilde  ($\widetilde{\cdots}$) for quantities in the SRF.

\begin{table*}
\begin{minipage}{12cm}
\caption{Fluid and particle quantities in URF, DRF and SRF 
for an ultra-relativistic shock with $\Gs \gg 1$, and particles with
$\gamma \gg \Gs$, $\overline{\gamma} \gg 1$.}
\begin{tabular}{lccc} \hline
{\bf Quantity} & {\bf URF} & {\bf DRF} & {\bf SRF} \\
& & & \\
Shock speed & ${\displaystyle \bs \approx 1 - \frac{1}{2 \Gs^{2}}}
$ & $\bsd \approx \third$ & $0$ \\
& & & \\
Upstream fluid speed & $0$ & 
${\displaystyle \overline{\beta}_{\rm u} \approx 
- \left(1 - \frac{1}{\Gs^{2}} \right)}$ 
& $\tbu = - \bs$ \\
& & & \\
Downstream fluid speed & ${\displaystyle \brel \approx 1 - \frac{1}{\Gs^{2}}}$ 
& 0 & $\tbd \approx - \third$ \\
& & & \\
Particle energy & 
${\displaystyle E = \Grel \: \left(1 + \brel \barm \right) \: \barE}$ & 
${\displaystyle \barE = \Grel \: \left(1 - \brel \mu \right) \: E}$ & $\widetilde{E}$ \\
& & & \\
Cosine flight direction & 
${\displaystyle \mu = \frac{\barm + \brel}{1 + \brel \barm}}$ & 
${\displaystyle \barm = \frac{\mu - \brel}{1 - \brel \mu}}$ &
$\widetilde{\mu}$\\ 
& & & \\
Edge of loss cone &
${\displaystyle \sin \theta_{\rm c} = 1/\Gs}$ &
${\displaystyle \barm = \bsd \approx \third}$ &
${\displaystyle \widetilde{\mu} = 0}$\\
& & & \\ \hline
\end{tabular}
\end{minipage}
\end{table*}

We will assume that the shock
has a speed $\bs$ in the URF, moving in the $z$-direction towards positive 
$z$. For an ultra-relativistic shock with $\Gs = (1 - \bs^{2})^{-1/2} \gg 1$ we
can use the expansion
\begin{equation}
\label{Gexp}
	\bs \approx 1 - \frac{1}{2 \Gs^{2}} + {\cal O}(\Gs^{-4}) \; .
\end{equation}
If in the observer's frame the fluid moves with speed $\bu$ upstream, and
a speed $\bd$ downstream, the velocity relevant for the shock acceleration 
process is the relative velocity between the up- and downstream gas which
follows from the relativistic velocity addition law:
\begin{equation}
\label{relvelo}
	\brel = \frac{\bu - \bd}{1 - \bu \bd} \; .
\end{equation}
This assumes that the scattering agent is passively advected by the flow on
both sides of the shock.
The velocity $\brel$ is the velocity of the downstream fluid seen from the URF,
or of the upstream fluid seen from the DRF.
	
For an ultra-relativistic shock with $\Gs \gg 1$, where
the magnetic field is dynamically unimportant, the jump condition relating the
up- and downstream velocities in the SRF quickly approaches the ultra-relativistic
jump condition 
\begin{equation}
\label{ERRH}
	\tbd \: \longrightarrow \: \third \; ,
\end{equation}
regardless of the equation of state of the gas on either side of the shock.
In this limit, the relative velocity of the up- and
downstream flow satisfies (e.g. 
Blandford \& McKee, 1976, see also Kirk \& Duffy, 1999)
\begin{equation}
\label{udgamma}
	\Grel \approx \frac{\Gs}{\sqrt{2}} \; \; , \; \; 
	\brel \approx 1 - \frac{1}{\Gs^{2}}  \; .
\end{equation}
This implies that in the DRF the shock moves with a velocity $\bsd \approx \third$
in the positive $z$-direction.

We will normally assume that the particles involved in the Fermi-type shock
acceleration process are relativistic in
all three reference frames, with particle Lorentz factors satisfying
$\gamma \gg \Gs$ and $\overline{\gamma} \; , \; \widetilde{\gamma} \gg 1$. 
The particle motion relative to the shock is determined by the shock speed
and the angle $\theta$ between the shock normal $\hatn$ and the particle momentum.
In the URF this angle is denoted by $\theta$ and its cosine,  
\begin{equation}
\label{angledef}
	\mu = \cos{\theta} \approx \beta_{\parallel}\; ,
\end{equation}
determines the component $\beta_{\parallel}$ of the particle's velocity $\bm{\beta}$ 
(in units of $c$) along the shock normal, for relativistic particles with $\beta \approx 1$.
Particle energy and direction cosine in the different frames are connected
by the well-known Lorentz transformations (e.g. Rybicki \& Lightman, 1979, Ch. 4),
here used in the limit $p \approx E$. Table 1 shows the relevant relations for 
transformations between the URF and DRF.

These relations allow one to derive different (but equivalent) expressions
for the energy gain which results from one shock crossing cycle.
Consider a particle going through a cycle where it crosses the shock from
upstream to downstream with energy $\Ei$ and direction cosine $\mud$, 
and re-crosses into the upstream medium, after elastic scattering in the downstream 
flow where $\barE$ remains constant, with an energy $\Ef$ and direction cosine  
$\muu$. In that case the energy ratio as measured by an upstream observer equals
\begin{equation}
\label{ratioup}
	\frac{\Ef}{\Ei} = \frac{1 - \brel \mud}{1 - \brel \muu} \; .
\end{equation}
One can also express the energy ratio in terms of the direction cosine $\bmuu$ 
measured by a downstream observer at the moment of re-entry into the upstream 
medium: 
\begin{equation}
\label{ratiomix}
 	\frac{\Ef}{\Ei} = \half \Gs^{2} \: \left( 1 - \brel \mud \right)
 	\left(1 + \brel \bmuu \right) \; .
\end{equation}
Here we have used $\Grel = \Gs/\sqrt{2}$.
This last expression is the basis for Vietri's (1995) claim that particles
can gain energy with $\Ef/\Ei \sim \Gs^{2}$ at each crossing cycle.

In the same way, a crossing cycle where the particle leaves the downstream
medium with direction cosine $\bmuu$ and re-enters with direction cosine
$\bmud$ after an elastic scattering or deflection upstream with constant $E$, 
will lead to a downstream energy ratio equal to
\begin{equation}
\label{ratiodwn}
	\frac{\bEf}{\bEi} = \frac{1 + \brel \bmuu}{1 + \brel \bmud} =
	\frac{1 - \brel \mud}{1 - \brel \muu}
	\; .
\end{equation}	
Comparing (\ref{ratioup}) and (\ref{ratiodwn}) shows explicitly that {\em on
average} the ratio of final and initial energy is the same in up- and
downstream rest frames.

\subsection{Initial shock encounter}

At the first encounter with the shock, all particles with $-1 < \beta \mu < \bs$ 
are overtaken by the shock. We first consider the simplest case, that of particles 
almost at rest in a cold upstream flow so that $\beta \ll 1$ and 
$\Ei \approx  m$ with $m$ the particle rest energy.
After the first shock crossing, the downstream energy of the particles is
$\barE = \Grel \: m$. These particles are immediately relativistic in the
DRF if $\Gs \gg 1$, as is assumed here. Those particles that are scattered back
into the upstream flow have an energy equal to (see Table 1)
\begin{equation}
\label{backfirst}
	\Ef \approx
	\half \Gs^{2} \: \left( 1 + \brel \bmuu \right) \: m
\end{equation}
Here we used $\Grel = \Gs/\sqrt{2}$. Since those particles returning 
upstream must have
$\third < \bmuu \le 1$, this implies that (initially) non-relativistic particles 
receive a boost in the upstream energy in the range
\begin{equation}
\label{firsttime}
	\mbox{$\frac{2}{3}$} \Gs^{2} < \frac{\Ef}{\Ei} \le \Gs^{2}
\end{equation}
at the end of their first crossing cycle.

If the upstream medium contains a pre-existing population of relativistic 
particles, distributed isotropically in the URF, 
one must average Eqn. (\ref{ratiomix}) 
over all direction cosines $-1 \le \mud < \bs$  with a weight proportional to the 
particle flux $\propto \bs - \mud$ across the shock into the downstream medium.
This yields an average initial boost equal to
\begin{equation}
\label{first2}
	\frac{\Ef}{\Ei} = \mbox{$\frac{2}{3}$} \Gs^{2} \:
	\left( 1 + \brel \bmuu \right) \; ,	 
\end{equation}
which lies in the range
\begin{equation}
\label{first3}
	\mbox{$\frac{8}{9}$}  \Gs^{2} < \frac{\Ef}{\Ei} \le 
	\mbox{$\frac{4}{3}$} \Gs^{2} \; .
\end{equation}		 
Here we assume that $\bmuu$ is uncorrelated with $\mud$.

We conclude that in all cases the energy gain at the {\em first} shock encounter
is of order $\Ef/\Ei \sim \Gs^{2}$ for those particles which return to the
upstream medium, so that Vietri's (1995) suggestion applies at the first shock
encounter. This provides a natural injection process for further particle 
acceleration by repeated shock crossings at ultra-relativistic shocks.
We return to this point in Section 6.3.

This conclusion can also be reached by considering the Rankine-Hugoniot relations	 
for an  ultra-relativistic shock (e.g. Blandford \& McKee, 1976) since, 
from a microscopic point of view, the situation is similar: in a shock
the kinetic energy of the upstream flow must be `randomized' so that
a hot downstream state arises behind the shock. 
The proper downstream energy density $\overline{e}$, pressure $\overline{P}$
and number density $\overline{n}$ of the downstream flow are related
to the corresponding upstream quantities by
\[
	\overline{e} \approx 3\overline{P} \approx 2 \Gs^{2} \: \left(
	e + P \right) \; \; \; , \; \; \; \overline{n} 
	\approx 2 \sqrt{2} \: \Gs \: n
	\; .
\]
The typical `thermal' energy per particle in the downstream is
\[
	\bar{E}_{\rm th} = \frac{\overline{e}}{\overline{n}} \approx
	\half \sqrt{2} \: \Gs \: \left( \frac{e + P}{n} \right) \; .
\]
In the case of a cold upstream medium ($P \ll e \approx nm$) one finds
$\bar{E}_{\rm th} \approx \Gamma_{\rm rel} m$, with $\Grel \approx \Gs/\sqrt{2}$
for an ultra-relativistic shock,
while in the case of a relativistically hot upstream fluid 
($P = e/3$) one finds
$\overline{E}_{\rm th} = \frac{4}{3} \: \Gamma_{\rm rel} \: E$, with $E = e/n$ 
the mean upstream energy per particle. 
A Lorentz-transformation back to the URF adds another factor $\sim \Gs$, 
so an upstream observer assigns these particles in both cases
an energy $\Ef \approx \Gs^{2} \: \Ei$ with $\Ei$ the initial upstream energy.

\subsection{Kinematical constraints on the energy gain at relativistic shocks}

We now consider true Fermi-type acceleration at an ultra-relativistic
shock of particles that have been downstream at least once.
Particles re-crossing the shock into the upstream region satisfy at the moment
of shock crossing 
\begin{equation}
\label{constraint}
	\barmu > \overline{\bs} \approx \third \; \; \; , \; \; \; 
	\mu > \bs\; 
\end{equation}
for the down- and upstream particle flight direction respectively.
These relations define the downstream {\em loss cone} with a
corresponding opening angle $\theta_{\rm c}$ around the shock normal in 
the upstream frame.  This opening angle follows from
\[
	\sin \theta_{\rm c} = \sqrt{1 - \bs^{2}} = \frac{1}{\Gs} \; .
\]	 
As $\theta_{\rm c}$ is small for $\Gs \gg 1$ we can use $\sin \theta_{\rm c}
\approx \theta_{\rm c}$ to write  
the condition (\ref{constraint}) for crossing the shock into the
upstream flow in terms of upstream variables  as
\begin{equation}
\label{losscone}
	\thu <  \theta_{\rm c} \approx \frac{1}{\Gs} \; .
\end{equation}
Particles that have just entered the upstream region reside within 
this loss cone. Upstream deflection or scattering must change the 
upstream flight angle to a value $\theta > \theta_{\rm c}$ before a 
new shock crossing cycle can begin.

As we will show below, all plausible scattering and deflection mechanisms are
only able to change the angle $\theta$ by an amount 
$| \Delta \theta | \sim  \theta_{\rm c}$ before particles are
again overtaken by the shock. As a result,
the upstream angular distribution of these particles is confined to a cone 
with opening angle $\Delta \theta \sim 2/\Gs \ll 1$ around the shock normal: 
an effect of relativistic beaming.
For ultra-relativistic shocks the small angle approximation, 
$\sin \theta \approx \theta$, is an excellent one 
for {\em all} particles 
in the upstream region which have interacted with the shock at least once, 
in particular those about to be overtaken by the shock with 
$1 < \Gs \thd \lesim 2$. 

Consider a particle performing a crossing cycle from upstream to 
downstream and back. The angles in this crossing cycle satisfy
\begin{equation}
\label{anglimits}
	\thu < \frac{1}{\Gs} < \thd \; .
\end{equation}	
If we use the small angle approximation $\mu = 1 - \half \theta^{2}$ in expression
(\ref{ratioup}) for the energy ratio in the URF at the end of the cycle, 
together with expansion (\ref{udgamma}) for $\brel$, one finds
(Gallant \& Achterberg, 1999):
\begin{equation}
\label{smallgain}
	\frac{\Ef}{\Ei} \approx \frac{2 + \Gs^{2} \thd^{2}}{2 + \Gs^{2} \thu^{2}}
	\; . 
\end{equation} 		
Because of the inequality (\ref{anglimits}) this ratio is always larger than unity.
But on average (as we will demonstrate explicitly
in our simulations) its value is
never much larger than $\Ef/\Ei \sim 2$ since both the
angular factors in this expression remain of order unity.
A similar conclusion holds for the energy ratio $\bEf/\bEi$
seen by a downstream observer in a cycle where the particle moves
from downstream to upstream and back, as is readily seen from Eqn. (\ref{ratiodwn}).

This shows that Vietri's (1995) suggestion does not
apply to these particles. If one uses expression (\ref{ratiomix}) for the
energy ratio in a cycle starting in the 
upstream region,  the factor
\begin{equation}
\label{Doppfact}
	1 - \brel \mud \approx \frac{\thd^{2}}{2} + \frac{1}{\Gs^{2}} 
\end{equation} 
is always of order $\Gs^{-2}$ due to relativistic beaming, while the factor 
$\frac{4}{3} < (1 + \brel \bmuu) \le 2$ is always of order unity. Once again
one concludes that the value of $\Ef/\Ei$ must be of order unity.

\section{Deflection and scattering}

As stated above, the process of diffusive shock acceleration in the non-relativistic
case relies on scattering on magnetic irregularities to confine the accelerated
particles near the shock for some time. For the case of ultra-relativistic shocks we 
will consider two cases in some detail: the case of regular deflection upstream and
strong scattering downstream, and the case where strong scattering operates on 
both sides of the shock.

\subsection{Regular deflection}
 
 The simplest way upstream particles will leave the loss cone is
 through deflection by the Lorentz force in a uniform upstream magnetic field. 
 Without loss of generality we will choose the shock normal in the $z$-direction, 
 while the magnetic field is assumed to lie in the $x-z$ plane:  
 $\bm{B} = (B_{\perp} \: , \: 0 \: , \: B_{\parallel})$.

 The equation of motion for an ultra-relativistic particle of charge $q = Ze$ 
 and energy $E$ with gyration frequency $\Omega_{\rm g} = ZeBc/E$ 
 in a magnetic field $\bm{B} = B \: \hatb$ reads 
 \begin{equation}
 \label{Lorentz}
        \frac{{\rm d} \bbet}{{\rm d} t} = 
        \Omega_{\rm g} \left( \bbet \btimes \hatb \right) \; .
 \end{equation}
This equation can be solved approximately using the fact that
 \[
        \bpar = 1 - {\cal O}(\Gs^{-2}) \; \; \; , \; \; \; 
        \left| \bperp \right| \approx \theta = {\cal O}(\Gs^{-1}) \ll 1  
 \]
 at {\em all} times when the particle is upstream.
 Here $\bpar \simeq \mu$ is the component of the particle velocity
 along the shock normal, and 
 $\bperp = (\beta_{x} \: , \: \beta_{y} \: , 0)$
 the velocity component in the plane of the shock, all measured in the 
 upstream frame.
 Unless the field is almost along the shock normal so that
 \begin{equation}
 \label{inclination}
        \epsilon \equiv \left| \frac{\Bpll}{\Gs \: \Bprp} \right|
        \gesim 1 \; ,
 \end{equation}
 one can solve (\ref{Lorentz}) by iteration, with only
 $\Omega_{\perp} \equiv ZeB_{\perp}c/E$ appearing in the equations to leading
 order in $\epsilon$. We only consider the case $\epsilon \ll 1$, 
 covering almost all possible field orientations except for a fraction
 $\sim 1/\Gs^{2}$ of the solid angle. We assume that the particles
 are ultra-relativistic so that we can put 
 $\beta = \sqrt{\beta_{\parallel}^{2} + \beta_{\perp}^{2}} = 1$,
 and choose $ZeB_{\perp} \ge 0$ without loss of generality.   

Under these assumptions one finds for a particle that crosses the shock at $t = 0$
and $z = 0$ with initial momentum direction corresponding to 
$\bm{\beta}(0) = (\beta_{x \rm{i}} \: , \: \beta_{y \rm{i}} \: , \: \beta_{z {\rm i}})$:
\begin{eqnarray}
	\beta_{x}(t) & = &  \beta_{x \rm{i}} \; ,\nonumber \\
	& & \nonumber \\
\label{orbit}	
	\beta_{y}(t) & \approx & \beta_{y \rm{i}} + \Omega_{\perp}t \; , \\
	& & \nonumber \\
	z(t) & \approx & \beta_{z {\rm i}}t - 
	\half \: \beta_{y \rm{i}} \left( \Omega_{\perp} t \right)^{2}
	- \mbox{$\frac{1}{6}$} \: \left( \Omega_{\perp} t \right)^{3} 
	\nonumber \; .
\end{eqnarray}
\nskip
The expression for
$z(t) = \int_{0}^{t} {\rm d} t' \: \beta_{z}(t')$  follows from 
$\beta_{z} \approx 1 - \half \beta_{\perp}^{2}$.	
We have consistently neglected all terms of order $1/\Gs$ and $\epsilon$
with respect to unity in the expressions for $\beta_{x}$ and $\beta_{y}$. 
The time $t_{\rm u}$ of the next encounter with the shock, 
the upstream residence time, is obtained by putting 
$z(t_{\rm u}) = z_{\rm s} = \bs t_{\rm u}$, which yields:
\nskip
\begin{equation}
\label{enctime}
	\Omega_{\perp} t_{\rm u} = 
	\half \left[ 9 \beta_{y \rm{i}}^{2} + 12 \left( \frac{1}{\Gs^{2}} - \beta_{\perp 0}^{2}
	\right) \: \right]^{1/2} - \mbox{$\frac{3}{2}$} \beta_{y \rm{i}} \; .
\end{equation}
\nskip
\noindent
The corresponding values of $\beta_{y}(t_{\rm u}) \equiv \beta_{y {\rm f}}$ and 
the position $z_{\rm s}$ can be found from substituting this expression in 
Eqn. (\ref{orbit}). 
In particular we find $\beta_{x {\rm f}} = \beta_{x {\rm i}}$ and
\begin{equation}
\label{ydefl}
	\beta_{y \rm f} = - \frac{\beta_{y {\rm i}}}{2} +
	\sqrt{\displaystyle \frac{3}{\Gs^{2}} - 3 \beta_{x {\rm i}}^{2} -
	\frac{3}{4} \: \beta_{y {\rm i}}^{2}} \; ,
\end{equation}
c.f. Gallant \& Achterberg (1999).
The angle with respect to the shock normal follows from
\begin{equation}
\label{angle}
	\beta_{x}^{2} + \beta_{y}^{2} = \sin^{2}{\theta} \approx \theta^{2} \; .
\end{equation}		 
In figure 1 we give a graphical representation of the position of the
particle in the $\beta_{x}$-$\beta_{y}$ plane when it is overtaken by the 
shock at $t = t_{\rm u}$, as a function of the initial conditions 
($\beta_{x {\rm i}}$, $\beta_{y {\rm i}}$) at the moment the particle enters 
the upstream region.  
  
\begin{figure}
\centerline{\psfig{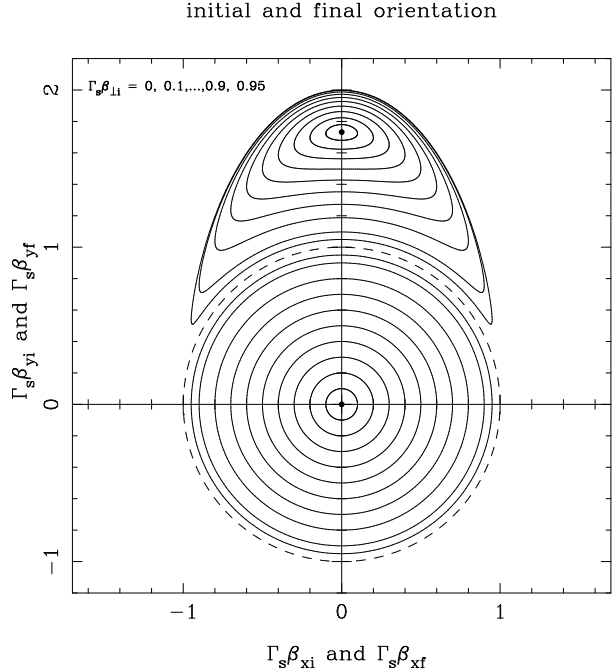}}
\caption{The location of a particle in the $\Gs \beta_{x}$-$\Gs \beta_{y}$ 
plane when it is overtaken by the shock as a 
function of its location when it enters the upstream region. 
Shown is the location of particles entering on concentric circles within the
loss cone centered on the shock normal. The dashed circle with radius 
$\Gs \beta_{\perp} = 1$ corresponds to the edge of the loss cone. 
The concentric circles have radii corresponding to 
$\Gs \beta_{\perp {\rm i}}$ equal to 0, 0.1, 0.2, $\cdots$, 0.9 and 0.95 
(the dot and the thick solid circles). These circles map onto the
kidney-shaped curves along lines of constant $\beta_{x}$. 
The origin $\beta_{x {\rm i}} = \beta_{y {\rm i}} = 0$,
corresponding to a particle entering the upstream flow along the shock normal, 
maps to $\beta_{x {\rm f}} = 0$, $\Gs \beta_{y {\rm f}} = \sqrt{3}$.
This is indicated by the thick dots.
The larger changes in the angle $\theta \approx | \bm{\beta}_{\perp}|$
occurs for particles with $\beta_{y {\rm i}} < 0$ and
$\beta_{x {\rm i}}= 0$, with the largest change 
ocurring for $\Gs \beta_{y \rm{i}} = -1$. 
These particles must turn through most of the loss cone before they 
leave it and are overtaken by the shock.
In contrast, particles with $\beta_{y {\rm i}} > 0$ change their orientation by
a relatively small amount, leaving the loss cone almost immediately. 
This figure is for $ZeB_{\perp} > 0$.
If the perpendicular field has the opposite sign, this figure should be mirrored
with respect to the $\beta_{x}$-axis.} 
\end{figure}

The largest deflections for given $\beta_{y {\rm i}}$ 
occur when the particles enter the upstream region with
$\beta_{x {\rm i}} = 0$. In that case we have in the small 
angle approximation
\[
	\thu \approx |\beta_{y {\rm i}}| \; \; \; , \; \; \;
	\thd \approx |\beta_{y {\rm f}}| \; ,
\] 
and particle trajectories remain in the $y-z$ plane.
In figure 2 we give the value of $\beta_{y {\rm f}} = \thd$, the energy gain and 
the upstream residence time as $\Gs \Omega_{\perp} t_{\rm u}$. 
The largest value for $\thd$ occurs when 
$\Gs \:\beta_{y {\rm i}} = -1$: then $\Gs \: \thd = 2$. However, the largest 
energy change from Eqn. (10), 
corresponding to $\bEf/\bEi \approx 2.62$, occurs for 
$\Gs \beta_{y \rm{i}} \approx -0.27$.    

\begin{figure}
\centerline{\psfig{figure=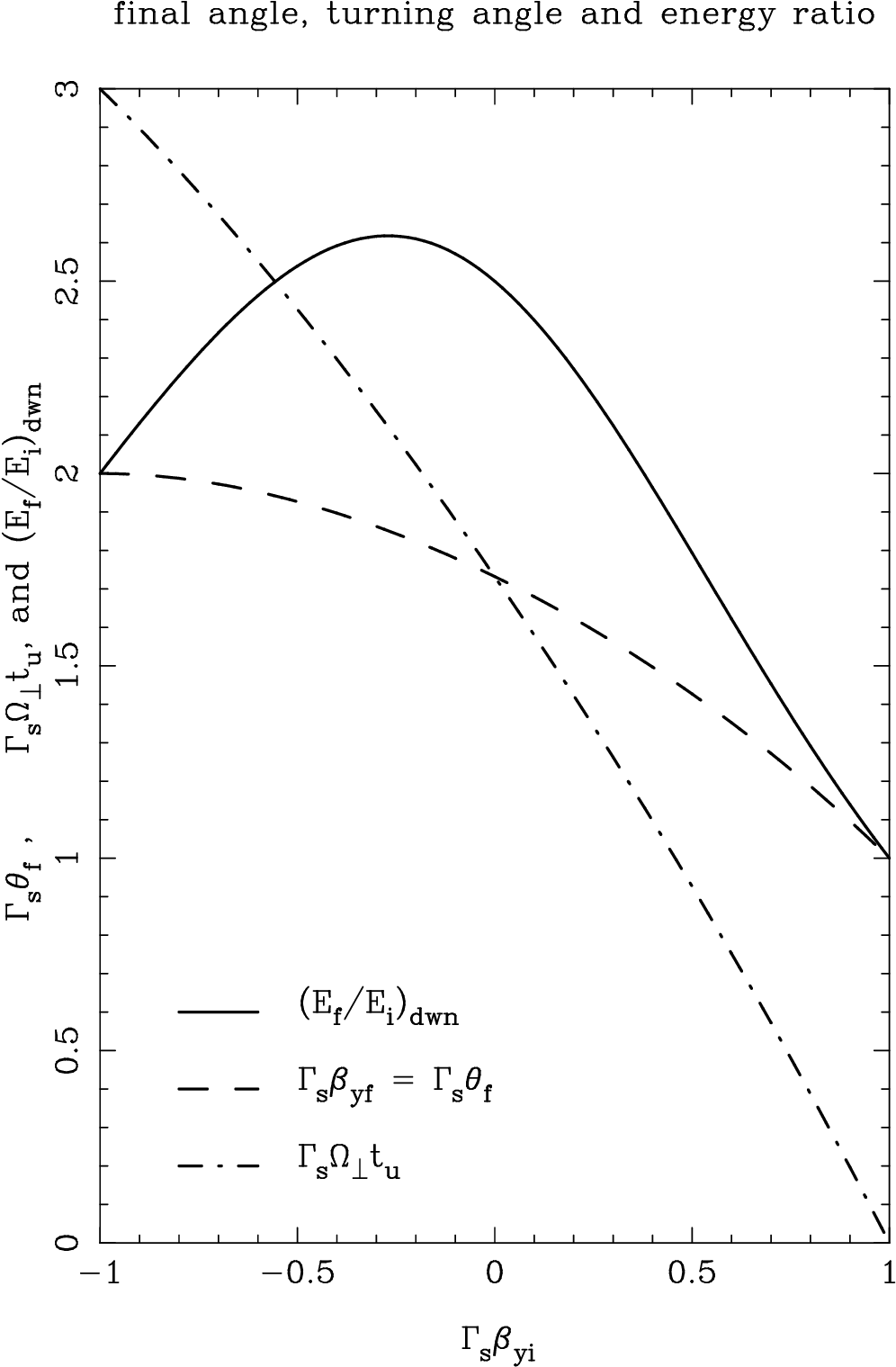,width=8cm,angle=0}}
\caption{The final angle $\thd \approx \beta_{y {\rm f}}$, the upstream
residence time, plotted as  $\Gs \Omega_{\perp}t_{\rm u}$,
and the downstream energy gain $\bEf/\bEi \equiv (\Ef/\Ei)_{\rm dwn}$ in the
downstream-upstream-downstream cycle starting with $\thu = |\beta_{y {\rm i}}|$
and ending with $\thd = \beta_{y {\rm f}}$, all as a function of
$\beta_{y {\rm i}}$ with respect to the shock normal at the moment
the particle enters the upstream flow. Curves are shown
for those particles moving in the
$y$-$z$ plane, i.e. $\beta_{x {\rm i}} = \beta_{x {\rm f}} = 0$.}
\end{figure}

These results set the following limits on the angle $\thd$ and its downstream
equivalent $\bthd = \cos^{-1}(\bmud)$ with respect to the shock normal when particles 
enter the downstream medium after regular deflection by an upstream magnetic field:
\begin{equation}
\label{anglimits2}
	1 < \Gs \thd < 2 \; \; \; \Longleftrightarrow \; \; \; 
	\third > \bmud > - \third \; .
\end{equation}	

\subsection{Scattering}

We now consider the circumstances under which scattering rather than 
deflection by an ambient upstream field changes the upstream particle momentum.
In the case of ultra-relativistic shocks the small opening angle
$\theta_{\rm c} \sim 1/\Gs$ of the loss cone strongly limits the regime in
which the direction of the upstream particle momentum can exhibit diffusive 
behaviour.

As a simple model for scattering of a charged particle by waves or random
magnetic fields consider a process where the angle between shock normal
and particle momentum changes at a constant rate, whose sign is chosen at
random, during a correlation time 
$t_{\rm c}$ , 
\begin{equation}
\label{regular}
	\theta(t + \Delta t) = \theta(t) \pm 
	\theta_{\rm rms} \: \left(\frac{\Delta t}{t_{\rm c}} \right)
	\; \; \; \mbox{for $\Delta t \le t_{\rm c}$.}
\end{equation}
For $t \gg t_{\rm c}$ this process corresponds to a random walk in 
$\theta$ with step size $\theta_{\rm rms}$ and corresponding diffusion 
coefficient
\begin{equation}
\label{diffdef}
	{\cal D}_{\theta} =
	 \frac{\theta_{\rm rms}^{2}}{2 t_{\rm c}} \; .
\end{equation}
An upstream particle will only enter the regime where the momentum direction
diffuses if it is not swept up by the shock within a correlation time $t_{\rm c}$.
Since the regular motion within a correlation time as decribed by (\ref{regular})
is formally equivalent with the case considered in the previous Section with
$\beta_{x} = 0$, $|\beta_{y}| = \theta$ and $\Omega_{\perp} \: \longrightarrow
\theta_{\rm rms}/t_{\rm c}$, the capture time $t_{\rm u}$ in the regular regime 
for a particle crossing the shock with an angle $0 \le \thu < 1/\Gs$ follows from
(\ref{enctime}) as
\begin{equation}
\label{regcapture}
	\theta_{\rm rms} \: \left( \frac{t_{\rm u}}{t_{\rm c}} \right) =
	\sqrt{ \frac{3}{\Gs^{2}} - \frac{3}{4} \: \thu^{2}} \mp
	\frac{3}{2} \: \thu  \le \frac{3}{\Gs} \; .
\end{equation}
where the minus (plus) sign applies in the half space $\beta_{y {\rm i}} = \thu$
and $\beta_{y {\rm i}} = - \thu$ respectively.
The maximum value of $t_{\rm u}$ occurs for $\Gs \beta_{y {\rm i}} = -1$ 
(see figure 2).	
This equation has a solution with $t_{\rm u} < t_{\rm c}$ for all allowed
values of $\thu$ provided
\begin{equation}
\label{steplim}
	\theta_{\rm rms} > \frac{3}{\Gs} = 3 \theta_{\rm c} \; .
\end{equation}
If this condition on the stepsize of the random walk in $\theta$ is satisfied,
upstream particles are overtaken by the shock before the direction of their 
momentum starts to diffuse. In that case, all the results for regular 
deflection apply.	

If condition (\ref{steplim}) is not satisfied, diffusion will start before
the particle is overtaken. 
The {\em average} position at time $t$ of particles entering the upstream medium
at $t = 0$ at an angle $\thu$ with respect to the shock normal can be calculated
as (see Appendix B of Achterberg \etal (1999)  using units with $c = 1$)
\begin{equation}
\label{avepos}
	\left< z(t) \right> =
	\left(1 - \half \thu^{2} \right)t - {\cal D}_{\theta}t^{2} \; ,
\end{equation}
assuming that ${\cal D}_{\theta}t \ll 1$ but $t \gg t_{\rm c}$.
The average value of $\theta^{2}$ satisfies in the same limit
\begin{equation}
\label{avang}
	\left< \theta^{2}(t) \right> = \thu^{2} + 4 {\cal D}_{\theta} t \; .
\end{equation}
The diffusive change in $\theta^{2}$ differs from the naive estimate,
$\left< \theta^{2}(t) \right> - \thu^{2} = 2 {\cal D}_{\theta}t$, 
due to the effect of dynamical friction. 
The typical time $t_{\rm u}$ at which a particle
is overtaken by the shock can be estimated by putting 
$z_{\rm s} = \beta_{\rm s} t_{\rm u} = \left< z(t_{\rm u}) \right>$.
This yields the relation
\begin{equation}
\label{captime}
	{\cal D}_{\theta}t_{\rm u} \sim \frac{1}{2 \Gs^{2}}
	\left( 1 - \Gs^{2} \thu^{2} \right) \; ,  	
\end{equation}
and allows us to estimate the angle at which the particle re-crosses into the
downstream medium as (Gallant \& Achterberg, 1999)
\begin{equation}
\label{capangle}
	\thd^{2} \sim \left< \theta^{2}(t_{\rm u}) \right> =
	\frac{2}{\Gs^{2}}  - \thu^{2} \; .
\end{equation}
These relations are only approximate since they neglect the correlation
between the shock re-crossing time and the re-crossing angle. Note that
they do satisfy the intuitive requirement that in this regime particles
entering the upstream region near the edge of the loss cone, $\thu \lesim 1/\Gs$, 
are recaptured almost immediately by the shock just outside
the loss cone with $\thd \gesim 1/\Gs$.

The typical downstream ratio of the final and initial energy in a 
crossing cycle where the particle moves from downstream to upstream and back
follows from Eqns. (\ref{ratiodwn}) and (\ref{capangle}) as
\begin{equation}
\label{meanEchange}
	\frac{\bEf}{\bEi} \approx
	\frac{4 - \Gs^{2} \thu^{2}}{2 + \Gs^{2} \thu^{2}}
\end{equation}
This downstream energy ratio falls steadily from $\bEf/\bEi = 2$ at
$\thu = 0$ to $\bEf/\bEi = 1$ at the edge of the loss cone where $\Gs \thu = 1$.	
Note however that this is an estimate based on averages, so that invidual
particles can have a larger energy gain.	

\subsection{Scattering agents: waves and random magnetic fields}

In diffusive acceleration near non-relativistic shocks one usually relies
on {\em gyro-resonant} scattering of the particles by low-frequency
MHD waves (Alfv\'en waves) supported by a nearly uniform
magnetic field (e.g. Wentzel, 1974). The magnetic field perturbation $\delta \bm{B}$
in these waves satisfies $\delta \bm{B} \bdot \bm{B} = 0$.
In this process particles interact mainly with those waves that satisfy 
the resonance condition
\[
	\omega(\bm{k}) - k_{\parallel} v_{\parallel} \pm \Omega_{\rm g} = 0 \; ,
\] 
where $\omega(\bm{k})$ is  the wave frequency and $\bm{k}$ is the wave vector, 
with $k_{\parallel}$ the component along the mean magnetic field.
For low-frequency waves this usually means that their wavelength must be of
order of the particle gyration radius.

In the rather extreme circumstances considered here, this process will only
be of importance if $\Bprp \ll \Bpll/\Gs$ so that deflection by the large-scale
field can be neglected to lowest order. 
Note that this is only the case for a small fraction $\ll 1/\Gs^{2}$ of possible
field orientations. We will consider the 
limiting case $\Bprp = 0$ and $\Bpll \equiv B$.
The resonant scattering process itself corresponds to a random walk in $\theta$
with
\[
	\theta_{\rm rms} \sim \left(\frac{\delta B}{B} \right) 
	\; \; , \; \; t_{\rm c} \sim 2 \pi / \Omega_{\rm g} \; ,
\]
with $\delta B$ the typical amplitude of the waves at resonant
wavelengths. If (\ref{steplim}) is satisfied, corresponding to a wave amplitude 
\begin{equation}
\label{amplim}
	\delta B > 3B/\Gs \; ,
\end{equation}
upstream particles will never enter into the diffusive regime. In that case
they spend less than a gyroperiod upstream, and the resonance between waves
and particle can not be achieved. In this regime the {\em non-resonant}
`sloshing motion' of the particles in the magnetic field of waves with
wavelength $\lambda \gesim \rg$, which has an
amplitude $\Delta \theta \approx \delta B/B$, is
sufficient to deflect them out of the loss cone so that they are overtaken.
This means that only for very small wave amplitudes does true diffusion
take place upstream, with a diffusion coefficient (Wentzel, 1974)
\begin{equation}
\label{padiffusion}
	{\cal D}_{\theta} \sim \Omega_{\rm g} \: 
	\left( \frac{\delta B}{B} \right)^{-2}_{\lambda \sim \rg} \; .
\end{equation}

If no large-scale upstream magnetic field is present, similar conclusions hold.
Consider a random field with rms amplitude $B_{\rm r}$ and coherence length $\lc$.
The simplest realization of this situation is a collection
of cells of size $2 \lc$ with a random orientation of the magnetic field $\bm{B}$
and a field strength $| \bm{B}| = B_{\rm r}$. 
The typical 
gyration radius of a particle in such a field is
\begin{equation}
\label{gyrorad}
	\rg(E) = \frac{E}{ZeB_{\rm r}} \; .
\end{equation}

Charged particles traversing such a random field exhibit a random walk 
in the momentum direction of the sort discussed in the previous Section, provided
$\lc \ll \rg(E)$. 
This random walk can be characterised by (Achterberg \etal 1999):
\begin{equation}
\label{randomw}
	\theta_{\rm rms} \sim \mbox{$\frac{4}{3}$} \:
	\frac{\lc}{\rg} \; \; \; , \; \; \; 
	t_{\rm c} \sim \frac{2 \lc}{c} \; .
\end{equation}		
The deflection of a particle in a single magnetic 
cell will be sufficient to lead to
a shock crossing if (\ref{steplim}) is satisfied, which in this case reduces to
\begin{equation}
\label{single1}
	\lc > \mbox{$\frac{9}{4}$} \: \frac{\rg(E)}{\Gs} \; .
\end{equation}
This condition is equivalent to
\begin{equation}
\label{single2}
	E < E_{\ast} \equiv \mbox{$\frac{4}{9}$} \:
	ZeB_{\rm r} \: \Gs \: \lc \; .
\end{equation}
Only particles with $E > E_{\ast}$ will be in the diffusive regime with
diffusion coefficient
\begin{equation}
\label{randiff}
	{\cal D}_{\theta} = \frac{c \lc}{3 \rg^{2}} \; .
\end{equation}
These calculations show that upstream, true diffusion of the particle
flight direction occurs only if the large-scale upstream magnetic field 
(if present) is closely aligned with the shock normal, and if the
deflection by waves or magnetic cells occurs in sufficiently small steps.
If this is not the case, particle sloshing motion in the waves or deflection
within a single magnetic cell is already
sufficient to deflect a particle out of the loss cone so that a new
shock crossing results.

In the downstream medium, on the other hand, particles have to turn through a
large angle of order $\Delta \overline{\theta} \sim 1$ in order to re-cross 
the shock, and strong diffusion in the angle $\overline{\theta}$ is needed.

\section{Cycle time and Maximum energy}

The maximum energy that can be achieved in Fermi-type acceleration at
an ultra-relativistic shock can be calculated in the same way as in the
non-relativistic case: acceleration proceeds until the energy gain per cycle
becomes equal to the energy losses incurred over a shock crossing cycle. 
Since the typical
energy gain per cycle at an ultra-relativistic shock satisfies
\[
	\Delta E = \Ef - \Ei \approx \Ei \; ,
\]
the acceleration time scale and the cycle time, which consists of the
time a particle resides in the up- and downstream region, 
$t_{\rm cy}(E) = t_{\rm u} + t_{\rm d}$,		
are essentially the same. Since the relative speed between the URF and
DRF is relativistic, one has to be careful about the frame used
in calculating the two contributions to $t_{\rm cy}$.
In particular, if a particle of energy $\barE$ in the DRF spends a
time $\overline{t}_{\rm d}$ in that frame before re-crossing the shock,
the corresponding upstream energy is $E \sim \Gs \: \barE$,
and the downstream residence time, separating two events at the shock, 
is equal to $t_{\rm d} \sim \Gs \: \overline{t}_{\rm d}$. 
This means that the cycle time
measured by an upstream observer can be written as
\begin{equation}
\label{upcycle}
	t_{\rm cy}(E) \sim t_{\rm u}(E) +
	\Gs \: \overline{t}_{\rm d}(E/\Gs) \; .
\end{equation}

\subsection{Cycle time for regular deflection}
	
Let us first consider the upstream residence time in the case of
deflection by a large-scale magnetic field. In that case the particle
momentum must typically turn through an angle $\Delta \theta \sim 1/ \Gs$
before the particle is overtaken once again by the shock. This implies
\begin{equation}
\label{uptime}
	t_{\rm u}(E) \sim \frac{\rg(E)}{\Gs c} = 
	\frac{E}{Ze\Bprp \: \Gs c} \; , 
\end{equation}
in accordance with the results of Section 4.1. 

In the downstream medium we must appeal to some form of scattering.
If the downstream field consists simply of the compressed large-scale
upstream field resulting from the shock jump conditions, where the
perpendicular field component is amplified according to
$\overline{B}_{\perp} \approx 2 \sqrt{2} \: \Gs B_{\perp}$ while the
parallel field component remains the same, 
the downstream magnetic field will be almost completely aligned with the 
shock surface.
In that case shock acceleration can only result if there is efficient
cross-field diffusion of particles (Jokipii, 1987; Achterberg \& Ball, 1994)
so that particles can catch up with the shock, which moves with a speed
$\sim c/3$ with respect to the downstream medium. 
This diffusion is presumably due to (shock-induced) wave turbulence.

This requirement explains why the simulations of Ballard and Heavens (1992),
who follow {\em exact} particle orbits in a random magnetic field, find
much steeper spectra for particles interacting with a relativistic shock
($\bs \approx 0.98$). Without a stochastic process which allows for a rapid
decorrelation between particle and magnetic field in the downstream
region, only particles which
are located on a field line with a favorable geometry can re-cross the
shock. This reduces the average return probability which leads to
a steepening of the spectrum.

As long as the turbulence is not so strong that the downstream field geometry
becomes completely chaotic, such a situation can only be realised if the 
cross-field diffusion coefficient $D_{\perp}$ is close to the maximum possible 
value: {\em Bohm diffusion} with diffusion coefficient 
$\overline{D}_{\rm B} = c \overline{\rg}/3$,
where particles randomly move across the field with a step size 
of order the particle gyration radius every gyration period. 
This follows from the requirement that the diffusive 
stepsize $\Delta \overline{z}$ along the shock normal, which downstream is 
almost perpendicular to the downstream field, 
must satisfy (Achterberg \& Ball, 1994)
\[
	\Delta \overline{z} > c \bsd \overline{\tau} 
	\approx c \overline{\tau} /3 \; , 
\]
with $\overline{\tau}$ the time interval between diffusive steps. For cross-field
diffusion  in the weak turbulence limit one has $\Delta \overline{z} \sim
\overline{\rg}$, with $\overline{\rg}$ the downstream gyroradius.
The above condition then corresponds to 
$\overline{\tau} < 3 \overline{\rg}/c$ and to a diffusion coefficient which 
must satisfy
\[
	\overline{D}_{\perp} \sim \frac{\overline{\rg}^{2}}{2 \overline{\tau}} 
	> \frac{c \overline{\rg}}{6} \equiv \half \: \overline{D}_{\rm B} \; .
\]
This is the case on which we will base our estimate below (Eqn. \ref{dwncyc})
for the downstream residence time.

If the downstream field is completely chaotic, leading to quasi-isotropic 
diffusion of the momentum direction downstream, particles in general will find
their way back to the shock, as our simulations show. We will 
show results for the typical downstream residence time for this case in 
Section 6.

The typical residence time of a particle in the downstream medium is
(Drury, 1983) 
\begin{equation}
\label{dwntime}
	\overline{t}_{\rm d}(\barE) \sim \frac{4 D_{\perp}}{c^{2} \bsd} 
	\sim \frac{4 \: \barE}{Ze \overline{B}_{\perp} c} \:
	\left( \frac{\overline{D}_{\perp}}{\overline{D}_{\rm B}} \right) \; .
\end{equation}
Simple shock compression leads to a downstream
magnetic field with $\overline{B}_{\perp} = 2 \sqrt 2 \: \Gs B_{\perp}$,
so scaling law (\ref{upcycle}) gives the corresponding time in the
URF for Bohm diffusion ($\overline{D}_{\perp} \approx \overline{D}_{\rm B}$) as
\begin{equation}
\label{dwncyc}
	t_{\rm d}(E) =  \Gs \: \overline{t}_{\rm d}(E/\Gs) =
	\frac{\sqrt{2} \: E}{Ze B_{\perp} \: \Gs c} \; .
\end{equation}
Comparing this with expression (\ref{uptime}) one sees that 
$t_{\rm d} \sim t_{\rm u}$ at a given energy. If the cross-field diffusion
is strongly enhanced by the effect of field-line wandering 
(Achterberg \& Ball, 1994) in a turbulent downstream flow, the downstream
residence time could be increased significantly so that the particle spends
most of its time downstream. On the other hand, if turbulence is strongly enhanced
downstream so that scattering proceeds at a rate with effective collision
frequency $\nu_{\rm c} = 1/t_{\rm c} \gg \Omega_{\rm g}$, particles are
no longer effectively tied to field lines. In that case particles will diffuse 
almost isotropically downstream, and will spend most of a shock 
crossing cycle in the upstream flow.

\subsection{Cycle time for upstream momentum diffusion}

If there is no large-scale field, but rather a randomly oriented field with 
coherence length $\lc$, and if $E \gesim ZeB_{\rm r} \Gs \lc$ so that
the momentum direction diffuses between shock crossings,
the upstream residence time is roughly the time it takes the momentum
direction to diffuse through an
angle $\sim 1/\Gs$ in the random field. Using (\ref{randiff}) one finds
\begin{equation}
\label{diffup}
	t_{\rm u}(E) \sim \frac{1}{\Gs^{2} \: {\cal D}_{\theta}} \approx
	\frac{3 E}{ZeB\: \Gs c}\: \left( \frac{\rg(E)}{\Gs \lc} \right) \; . 
\end{equation}
The extra factor with respect to the regular case is always larger than
unity because of condition $\lc \lesim \rg/\Gs$ for the diffusion regime
to apply upstream.

A similar calculation downstream, where particles must typically turn 
through an angle $\Delta \bar{\theta} \sim 1$, yields:
\begin{equation}
\label{diffdwn}
	\overline{t}_{\rm d}(\barE) \sim 
	\frac{3 \barE}{Ze \overline{B}c} \: 
	\left( \frac{ \overline{\rg}(\barE)}
	{\overline{\lc}} \right) \; .
\end{equation}		
The amount of additional field amplification downstream 
above shock compression can be parameterised by the ratio
\begin{equation}
\label{amppar}
	\xi_{\rm B} \equiv \frac{\overline{B}}{2 \sqrt 2 \: \Gs B_{\perp}}
	\ge 1 \; .
\end{equation}
Using
\[
	\overline{\rg}(\barE) \approx \rg(E)/(2 \sqrt 2 
	\: \Gs^{2} \xi_{\rm B}) \; ,
\]
one can write the corresponding time measured by an upstream observer as
\begin{equation}
\label{diffdwnu}
	t_{\rm d}(E) =  \Gs \: \overline{t}_{\rm d}(E/\Gs) \approx
	\frac{3 E}{ZeB_{\perp} \: \Gs c}\: 	 
	\left( \frac{\overline{\rg}(\barE)}{8 \xi_{\rm B} \overline{\lc}} 
	\right) \; .
\end{equation}
Downstream the particles will diffuse provided $\overline{\lc} < \overline{\rg}$.
If downstream field amplification is large, $\xi_{\rm B} \gg 1$, as
is often assumed in afterglow models for Gamma Ray Bursts, the downstream
residence time would typically be less than the upstream residence time. 

\subsection{Maximum energy estimates}

The above estimates for the cycle time allow one to calculate the maximum energy that can be
achieved in Fermi-type acceleration at an ultra-relativistic shock, generalizing the
results of Lagage and Cesarsky (1983), Drury (1983) and Heavens (1984) to the relativistic case. 
Consider a propagating spherical shock with radius $\Rs$. 
The finite age of the shock, $t_{\rm s} \approx \Rs/c$,
limits the energy of shock-accelerated particles through the requirement
$t_{\rm cy}(E) < \Rs/c$. The best possible case arises when 
$t_{\rm d} \ll t_{\rm u}$. 
For regular deflection upstream one then finds
\begin{equation}
\label{maxE1}
	E_{\rm max} \approx ZeB_{\perp} \: \Gs \Rs \; .
\end{equation}
If the particle momentum starts to diffuse in a random upstream field,
which requires $E \gesim E_{\ast} \sim Ze B_{\rm r} \: \Gs \lc$, the
maximum energy equals
\begin{equation}	
\label{maxE2}
	E_{\rm max} \approx Ze B_{\rm r} \: \Gs \: 
	\left(\Rs \lc/3 \right)^{1/2} \; .
\end{equation}
This expression for $\Emax$ is only relevant if $\Rs > \lc$. 

Similar limits on the maximum energy
apply to a standing shock in a spherical wind or in a jet with
constant opening angle. Expansion losses of relativistic
particles in a relativistic flow satisfy (e.g. Webb, 1985)
\begin{equation}
\label{relexploss}
	\left( \frac{{\rm d} E'}{{\rm d} t'} \right)_{\rm exp}
	 = - \frac{E'}{3} \: \left[
	\frac{\partial \Gamma}{\partial t} + \grad \bdot (\Gamma \bm{V}) 
	\: \right] \; .
\end{equation}
Here $E'$ and $t'$ are the particle energy and time measured in the
fluid rest frame, and $\bm{V}(\bm{x} \: , \: t)$ and $\Gamma(\bm{x} \: , \: t)$ 
are the fluid velocity and associated Lorentz factor in the observer's frame.
This result applies if frequent scattering keeps the particle distribution 
nearly isotropic in the fluid rest frame. In that case one can use
the Lorentz-invariance of the energy loss rate,
${\rm d} E' / {\rm d}t' = {\rm d}E/{\rm d}t$
(e.g. Rybicki \& Lightman, 1979, Ch. 4), together with $E \sim \Gamma E'$.
These relations then
yield the expansion losses of relativistic particles ($E \approx pc$)
in the observer's frame:
\begin{equation}
\label{relexplobs}
	\left( \frac{{\rm d} E}{{\rm d} t} \right)_{\rm exp}
	 \approx - \frac{E}{3 \Gamma} \: \left[
	\frac{\partial \Gamma}{\partial t} + \grad \bdot (\Gamma \bm{V}) 
	\: \right] \; .
\end{equation} 
Therefore, expansion losses in a flow with typical velocity $\Vs$ and size
$\Rs$ are of order	
\[
	\left( \frac{{\rm d} E}{{\rm d} t} \right)_{\rm exp} \sim -
	\left( \frac{\Vs}{\Rs} \right) \: E \; ,   
\]	
and involve the same dynamical timescale $t_{\rm s} \sim R_{\rm s}/V_{\rm s}$.
Particles will no longer have a net energy gain in a crossing
cycle if $t_{\rm cy}(E) \ge t_{\rm s}$.

The above estimates for $\Emax$ neglect radiation losses.
For electrons (or positrons)
these losses can limit the maximum energy to lower values.
This was discussed for non-relativistic shocks by Blandford(1977) and Heavens (1984).
The energy loss rate for an electron with $E \gg m_{\rm e}c^{2}$
scales as
\begin{equation}
\label{synchloss}
	\frac{{\rm d}E}{{\rm d}t} = - \mbox{$\frac{4}{3}$} \:
	\sigma_{\rm T} c U_{\rm t} \: 
	\left(\frac{E}{m_{\rm e} c^{2}} \right)^{2}
	\; .
\end{equation}
Here $\sigma_{\rm T}$ is the Thomson cross section and
\[
	U_{\rm t} \equiv U_{\rm rad} + \mbox{$\frac{3}{2}$} \: 
	\frac{B_{\perp}^{2}}{8 \pi} \; .
\]
The first term in $U_{\rm t}$ is  the energy density of the radiation
field (assumed to be isotropic) leading to inverse Compton losses. 
The second term corresponds to synchrotron losses in the ambient magnetic field. 
Since all upstream particles are closely aligned with the shock normal only 
the field component $B_{\perp}$ contributes to the synchrotron losses.

Consider a particle entering the upstream medium at $t = 0$ 
with an energy $\Ei$. Its energy decays due to radiation losses as
\begin{equation}
\label{Edecay}
	E(t) = \frac{\Ei}{1 + t/\tau_{\rm sy}} \; ,
\end{equation}
with 
\[
	\tau_{\rm sy} \equiv 
	\frac{3 m_{\rm e} c^{2}}{4 \sigma_{\rm T} c U_{\rm t}}
	\left(\frac{m_{\rm e} c^{2}}{\Ei} \right) 
\]
the loss time at energy $\Ei$.		
Assuming regular deflection upstream, the angle $\theta$ between
particle momentum and shock normal changes according to
\begin{equation}
\label{angchange}
	\frac{{\rm d} \theta}{{\rm d} t} =
	\frac{ZeB_{\perp}c}{E(t)} = \Omega_{\rm i} \: 
	\left( 1 + \frac{t}{\tau_{\rm sy}} \right) \; ,
\end{equation}
with $\Omega_{\rm i} = ZeB_{\perp}c/\Ei$. The time to turn through an
angle $\Delta \theta$ is
\[
	t = \tau_{\rm sy} \: \left(
	\sqrt{1 + \frac{2 \Delta \theta}
	{\Omega_{\rm i} \: \tau_{\rm sy}}} - 1
	\right)
\]
and the electron energy is reduced in that time to
\begin{equation}
\label{Ereduct}
	E = \frac{\Ei}
	{\displaystyle 
	\sqrt{1 + \frac{2 \Delta \theta}{\Omega_{\rm i} \: \tau_{\rm sy}}}} \; .	
\end{equation}
Since shock crossings typically double the
particle energy in each cycle,  and since $\Delta \theta \sim 1/\Gs$, 
upstream losses will lead to a net energy {\em loss} in a cycle if
\[
	\Omega_{i} \tau_{\rm sy} \gesim 1/\Gs
\]
which corresponds to
\begin{equation}
\label{Esyup}
	E \gesim E_{\rm su} \equiv m_{\rm e}c^{2} \:
	\left( 
	\frac{3 e B_{\perp} \Gs}{4 \sigma_{\rm T} U_{\rm t}} 
	\right)^{1/2} \; .
\end{equation}
Losses in the downstream leg of the crossing cycle will become important
if the downstream residence time and the synchrotron loss time
are roughly equal. Assuming Bohm diffusion (Eqn. \ref{dwntime}) this condition
reads
\[
	\frac{4 \: \barE}{Ze \overline{B}_{\perp} c} \sim
	\frac{3 m_{\rm e} c^{2}}{4 \sigma_{\rm T} c \overline{U}_{\rm t}}
	\left(\frac{m_{\rm e} c^{2}}{\barE} \right) \; ,		
\] 	
which corresponds to an energy
\[
	\bar{E} \approx \barE_{\rm sd} =
	m_{\rm e} c^{2} \: \left( 
	\frac{3 e \overline{B}_{\perp}}{16 \sigma_{\rm T} 
	\overline{U}_{\rm t}} 
	\right)^{1/2} 
\]
in the DRF. In the URF these particles have an energy
\begin{equation}
\label{Esydwn}
	E_{\rm sd} \sim \Gs \: \barE_{\rm sd} =
	\Gs \: m_{\rm e} c^{2} \: \left( 
	\frac{3 e \overline{B}_{\perp}}{16 \sigma_{\rm T} 
	\overline{U}_{\rm t}} 
	\right)^{1/2} \; .	
\end{equation}
If Compton losses can be neglected ($U_{\rm t} \approx 3B_{\perp}^{2}/16 \pi$)
one has
\[
	E_{\rm su} \approx m_{\rm e} c^{2} \: 
	\left( \frac{4 \pi e \Gs}{\sigma_{\rm T} B_{\perp}} \right)^{1/2}
	\; \; \; , \; \; \; 
	E_{\rm sd} \approx 0.3 \:\frac{E_{\rm su}}{\sqrt{\xi_{\rm B}}}
	\; .
\]	
Here we have used definition (\ref{amppar}) in the second estimate, where
the numerical constant is $1/(8 \sqrt{2})^{1/2}$.
This shows that for $\xi_{\rm B} \approx 1$, up- and downstream losses
are roughly equally important, but that the downstream losses will 
determine the maximum energy
if significant field amplification occurs in the downstream flow.	
If losses are important, the maximum energy of shock-accelerated particles
will be
\[
	E_{\rm max} = \min \left( E_{\rm su} \: , \: E_{\rm sd} \right) \; .
\]

\section{Simulations}

We have performed a number of numerical simulations of the acceleration
process using the following assumptions:
\begin{itemize}
\item	The deflection mechanism in the upstream region is either
	pure scattering, or pure deflection by a homogeneous upstream field
	$B_{\perp}$;
\item	In the downstream region there is strong (diffusive) scattering and
	no regular deflection;
\item	Up- and downstream fluid states are connected by the ultra-relativistic
	jump condition
	$\widetilde{\beta}_{\rm u} \widetilde{\beta}_{\rm d} = \third$;
\item	Radiation losses are neglected.	
\end{itemize}
Our simulations employ the method of Stochastic Differential Equations
introduced by It\^o
(e.g. Gardiner, 1983; {\O}ksendal, 1991; Achterberg \& Kr\"ulls, 1992).	
We follow particles in the shock rest frame, advancing
the particle position $\widetilde{z}$ along the shock normal
at each time step as
\begin{equation}
\label{posfollow}
	\Delta \widetilde{z} = \widetilde{\Gamma}_{\rm u,d} \left(
	n_{z} - \widetilde{\beta}_{\rm u,d} \right) \: \Delta s_{\rm u,d} \; .
\end{equation}
Here $n_{z}$ is the component of the unit vector $\hatn = \bm{p}/|\bm{p}|$ 
along the shock normal, as measured in the URF or DRF respectively, and
$\Delta s_{\rm u,d} \equiv c \: \Delta t_{\rm u,d}$ the path length
increase of the particle in the upstream or downstream fluid rest frame, 
which is taken to be a constant. 
The quantities $\widetilde{\Gamma}_{\rm u,d}$ and $\widetilde{\beta}_{\rm u,d}$ 
are the up- and downstream fluid Lorentz factor and fluid speed in units of $c$ 
in the SRF. This expression follows straightforwardly from the 
Lorentz transformation for position between the SRF and the URF or DRF.
The shock is located at $\widetilde{z} = 0$, and when a shock crossing is
detected the particle momentum and energy is transformed between the
URF and DRF using {\em exact} Lorentz transformations. Particles
are then propagated further, starting at the shock.

To describe scattering, the particle flight direction $\hatn$ in the
URF and DRF is advanced according to (e.g. Achterberg et al., 1999)
\begin{equation}
	\hatn(s + \Delta s) = 
	\sqrt{1 - \left| \Delta \hatn_{\rm st} \right|^{2}} 
	\:\hatn(s) +  \Delta \hatn_{\rm st} \; .
\end{equation} 
In this expression $\Delta \hatn_{\rm st}$ is the stochastic change
in the orientation of the flight direction which results from 
scattering by random fields or waves. Expressed in terms of
the corresponding angular diffusion coefficient ${\cal D}_{\theta}$, this
change in the flight direction is given by
\begin{equation}
\label{angstep} 
	\Delta \hatn_{\rm st} = \sqrt{2 {\cal D}_{\theta} \: (\Delta s/c)} \: 
	\left( \xi_{1} \evec{1} + \xi_{2} \evec{2} \right) \; ,
\end{equation}
where $\xi_{1}$ and $\xi_{2}$ are two independent unit Wiener processes, 
drawn at each step from a Gaussian distribution with unit dispersion so 
that they obey the simple rules 
$\left< \xi_{1} \right> = \left< \xi_{2} \right> = 0$, 
$\left< \xi_{1}^{2} \right> 
= \left< \xi_{2}^{2} \right>= 1$ and 
$\left< \xi_{1} \xi_{2} \right> = 0$, where the brackets indicate an average
over many independent steps.
The two (arbitrary but mutually orthogonal) unit vectors $\evec{1,2}$ 
are chosen in the plane perpendicular to $\hatn$ so that 
$\Delta \hatn_{\rm st} \bdot \hatn = 0$. Note that the norm
$\hatn \bdot \hatn = 1$ is preserved identically.
It can be shown that this gives
an excellent approximation for isotropic scattering 
(i.e. ${\cal D}_{\theta}$ independent of $\theta$) 
for $|\Delta \hatn_{\rm st}| \ll 1$. 

Upstream, the value of $\Delta s$ is chosen in such
a way that the typical diffusive step $\Delta \theta_{\rm st}$
in the flight direction satisfies
$\Gs \Delta \theta_{\rm st} \le 0.1$. The downstream value $\overline{\Delta s}$
is chosen in such
a way that it matches the upstream value after the appropriate Lorentz transformation,
until the particle has diffused a distance of a few mean free paths behind
the shock. Then $\overline{\Delta s}$ is typically chosen ten times
larger.

In the case of regular deflection by a homogeneous field in the upstream
medium the analytical solution (\ref{ydefl}) is used to calculate the
orientation angle $\thd$ of the particle momentum vector at the moment the
particle recrosses the shock, given the flight direction $\thu$ with which the 
particle enters the upstream region. 

By running many, statistically independent, realisations
of this prescription and recording the particle energy and momentum
direction at each shock encounter, 
one can construct the particle distribution in
momentum space at the shock which results from repeated shock encounters.
This corresponds to the situation where a steady-state is reached
in the shock frame.
 
We also use a particle splitting method, where the loss of
particles from the acceleration process due to the large value of ${\cal P}_{\rm esc}$
is compensated by the introduction of additional particles,
so that the effects of Poisson noise in the simulations is minimized.
Typically particles are split every time they complete five crossing cycles,
by keeping track of the number $n_{\rm du}$ of downstream to upstream crossings.
In this splitting method, a particle is split
in into $N$ copy particles, each with weight $W=1/N$. We typically use
$N = 25-100$. These copy particles start their evolution with the same
initial conditions ($n_{\rm du}$, momentum and position) as the parent particle.
Thereafter, all copy particles evolve independenty, according to the stochastic 
algorithm outlined above.

The splitting procedure has no noticable effect on the particle distribution.
This is most easily seen by looking at the distribution of shock crossings,
as this quantity determines the particle splitting. The distribution is a
featureless power law, with the slope set by the escape probability 
${\cal P}_{\rm esc}$ (see figure 3).  Neither the position of the splitting
boundaries in terms of the number of shock crossings, nor the number of
particles created at each boundary influences this result in any noticable way. 
The same holds for the energy distribution (figure 4) which, 
apart from the transient effect of the
injection conditions at the beginning of the simulation, is once again a featureless power 
law in energy.

\begin{figure}
\centerline{\psfig{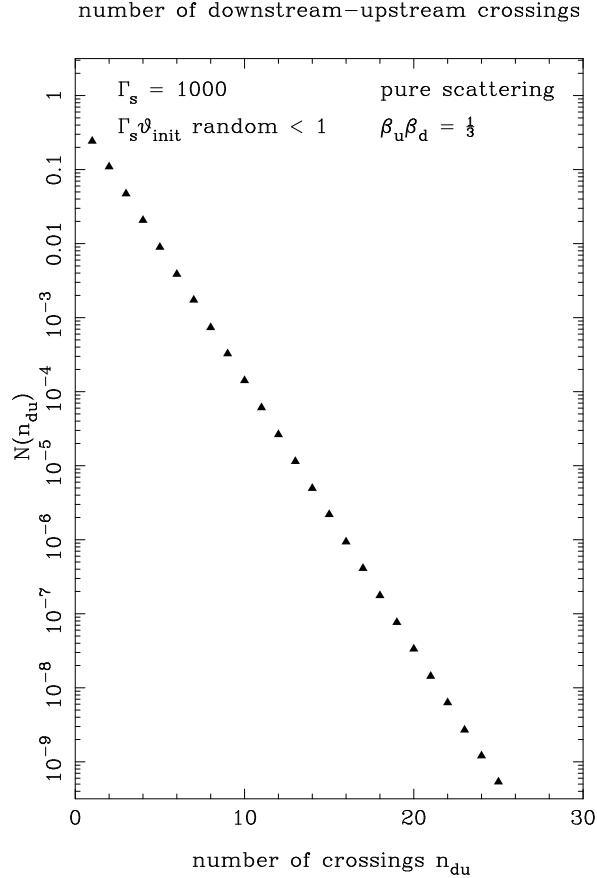}}
\caption{The distribution of the number of crossings $n_{\rm du}$ from the
downstream region into the upstream region. The distribution is a featureless power-law,
as should be the case for particles with an energy-independent return probability
$\Pret = 1 - \Pesc$. The distribution shows no sign of an influence of the particle splitting
employed to diminish the Poisson noise, which uses $n_{\rm du}$ in the splitting criterion.}
\end{figure}	

\begin{figure}
\centerline{\psfig{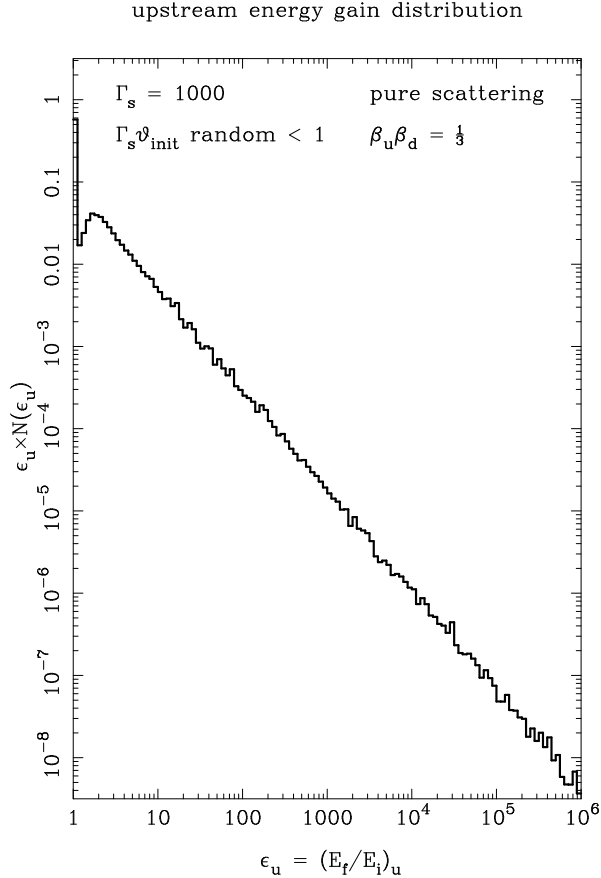}}
\caption{The distribution of particles as a function of the upstream energy gain $(\Ef/\Ei)_{\rm u}$,
for particles interacting with a shock with
$\Gs = 1000$. The particles are injected upstream at energy $\Ei$ with their flight direction distributed
randomly within the loss cone $\theta \le 1/\Gs$. A featureless power-law establishes itself
after a few crossings, signalling that the memory of the initial conditions has been erased. }
\end{figure}	

An absorbing boundary is placed sufficiently far downstream (at a distance
from the shock at least $10 \times$ the scattering mean free path) to remove 
particles from the simulation which have a vanishingly small chance of 
ever returning to the shock.  

\subsection{Case of pure scattering}

We first consider the case where particles are scattered on both sides of
the shock. For this case the semi-analytical results from Kirk \etal (2000) can
serve as a basis for comparison.
Table 2  gives the results for a number of simulations at different values
of $\Gs$. 

\begin{table}
\caption{The spectral slope $s$, the mean ratio of initial and final energy 
$<E_{\rm f}/E_{\rm i}>$ in a crossing cycle upstream $\longrightarrow$ downstream
$\longrightarrow$ upstream, 
and the return probability ${\cal P}_{\rm ret}$ for the case of strong scattering
on both sides of the shock.}
\label{scatscat}
\begin{tabular}{@{}lccc} \hline 
& & & \\
$\Gs$ & $s$ &  $<E_{\rm f}/E_{\rm i}>$ &  ${\cal P}_{\rm ret}$ \\
& & & \\
10 		& 2.230 $\pm$ 0.012 	& 1.97 $\pm$ 0.01 	& 0.435 $\pm$ 0.004 \\
10$^2$	& 2.219 $\pm$ 0.004	& 1.97 $\pm$ 0.01		& 0.439 $\pm$ 0.006 \\
10$^3$	& 2.213 $\pm$ 0.003	& 1.93 $\pm$ 0.02		& 0.437 $\pm$ 0.006 \\
\hline
\end{tabular}
\end{table}

The results presented in Sections 3 and 4 of this paper are the leading terms
in a formal expansion in powers of $1/\Gs^{2}$. This means that all the
simulations for $\Gs = 10, 100$ and $1000$ should give identical 
results for the slope of the distribution, the mean energy gain per cycle
and the return probability within about 1\%. Table 2 shows that this is indeed
the case.  

We can compare the slope $s$ of the energy distribution of the accelerated
particles as found in these  simulations with the
analytical results of Kirk \etal (2000), and the simulations of
Bednarz \& Ostrowski (1998). 
In our simulations $s$ is obtained from a least square fitting method
on the distribution of accelerated particles, at energies exceeding 
at least $30 \times$ the injection energy so that there is no `memory' of the
initial conditions at injection.   
The various values agree within quoted errors:

\[
	s = \left\{ \begin{array}{cl} 
	2.22 \pm 0.01 & \mbox{these simulations;} \\
	& \\
	2.2 & \mbox{Bednarz \& Ostrowski (1998);} \\
	& \\
	2.23 \pm 0.01 & \mbox{Kirk \etal (2000);} \\
	\end{array}
	\right.
\]
As a consistency check on the value of $s$ obtained from the
simulations, we can also use relation (\ref{Fermislope}) to
calculate it from the return probability $\Pret$
and from the mean energy gain per crossing $\left< \Ef/\Ei \right>$
obtained in these simulations. These are obtained by recording the number
of crossing cycles a particle completes, and the energy gain per
crossing.
Substituting the values from Table 2 we find
\[
	s = 1 + \frac{\ln(1/{\cal P}_{\rm ret})}{\ln \left< \Ef/\Ei \right>}
		= 2.23 \pm 0.02	
\] 
Figures 5 and 6 shows a more detailed comparison between the simulation results
and the semi-analytical results of Kirk \etal (2000).
	
\begin{figure}
\centerline{\psfig{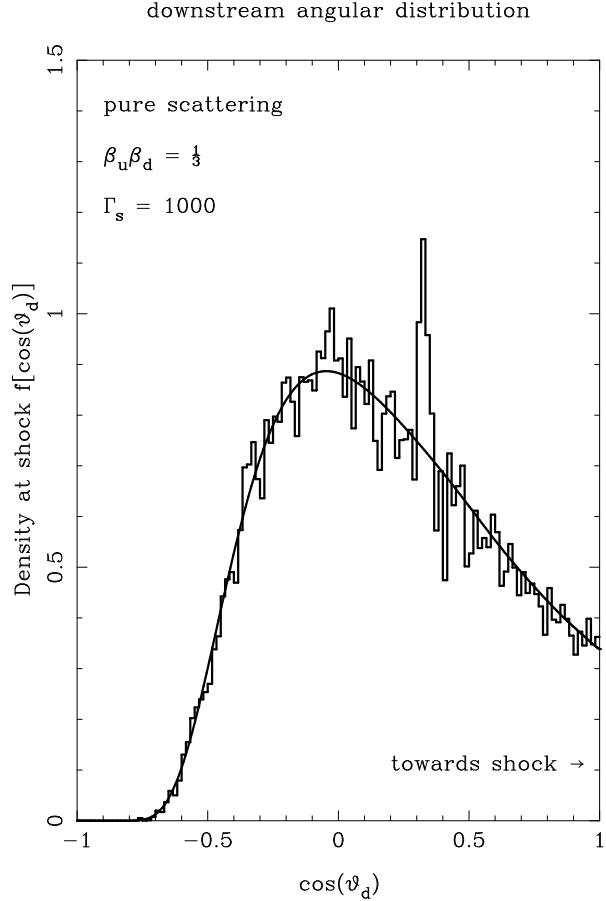}}
\caption{The angular distribution of particles at the shock in 
the downstream rest 
frame. The angle $\theta_{\rm d}$ is the angle between the
particle momentum and the shock normal, with $\theta_{\rm d} = 0$ corresponding
to a particle crossing the shock along this normal into the upstream medium. 
The distribution is normalized to a unit integral.
The histogram gives the results from our simulations, and the
smooth curve shows the semi-analytical result from the eigenfunction approach. 
The downstream loss cone corresponds to 
$\cos{\theta_{\rm d}} > \bsd \approx \third$.
This curve is for particles with an energy
$E > 30 \times E_{\rm inj}$, with $E_{\rm inj}$ the energy of a particle when
it starts true (Fermi-type) shock acceleration.
This ensures that the distribution has relaxed so that 
the initial conditions no longer influence the shape of the distribution.}
\end{figure}	

\begin{figure}
\centerline{\psfig{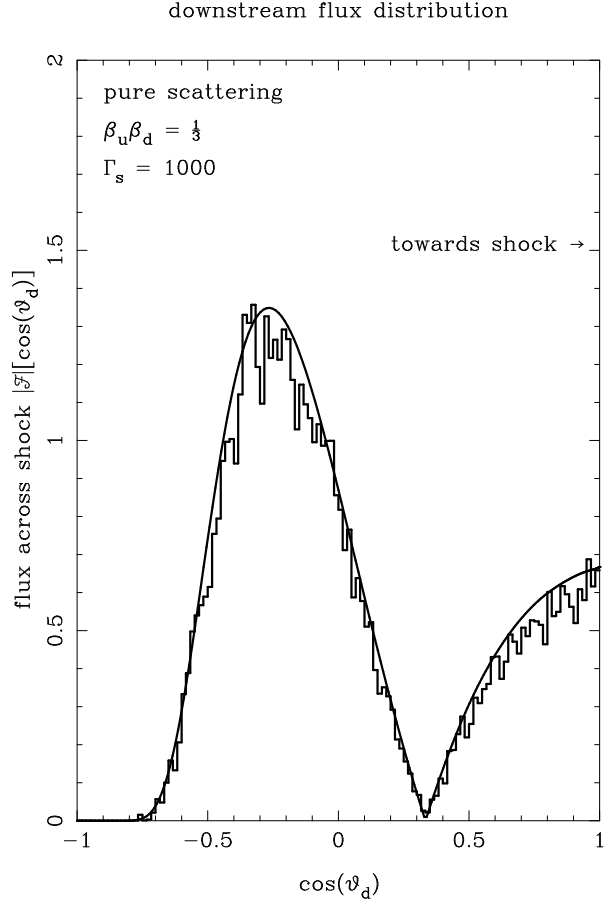}}
\caption{The distribution of the absolute value of the flux
${\cal F}$ across 
the shock as a function of $\cos(\theta_{\rm d})$ in downstream rest 
frame variables. 
The histogram gives the results from our simulations, and the
smooth curve is the semi-analytical result from the eigenfunction approach of
Kirk et al. (2000). The flux vanishes at the edge of the downstream 
loss cone, $\cos{\theta_{\rm d}} = \bsd \approx \third$, where
particles move exactly along the face of the shock.
This curve is for particles with an energy
$E > 30 \times E_{\rm inj}$, where the initial conditions no longer
influence the distribution.}
\end{figure}

Figure 5 shows the angular distribution of particles at the shock as
a function of the cosine of the angle between particle momentum and
shock normal, and Figure 6 shows the flux distribution. At the position
of the edge of the loss cone ($\cos(\overline{\theta}_{\rm d}) = \third$) 
there is an unphysical spike in the angular distribution. 
This spike is a discretization error, 
resulting from the fact that particles take discrete steps in angle, c.f. 
Eqn. (\ref{angstep}). This allows particles to be `trapped' near the edge of 
the loss cone since they can re-cross the shock after only one step.
In that limit, scattering is not well-represented by our method.
This angular distribution is in fact reconstructed from the particle {\em flux}
across the shock, which is the quantity recorded in our simulations.

We feel confident that this peak does not influence the results for the
slope $s$ of the spectrum of accelerated particles. The 
relevant parameters for the acceleration of particles, such as the mean energy
ratio $\left< \Ef/\Ei \right>$ and the return probability $\Pret$, are in fact
{\em flux} averages. Particles near the edge of the loss cone contribute a
negligible flux, and the spike is absent from the flux distribution of
Figure 6. The flux distribution obtained from the simulations agrees well
with the analytical results using the eigenfunction method described in 
Kirk \etal (2000).

Figure 7 shows the distribution of the upstream residence time $t_{\rm u}(E)$ and
the downstream residence time $\overline{t}_{\rm d}$ for $\Gs = 1000$,
both measured in the respective fluid rest frames for particles with
$E > 30 \times E_{\rm inj}$.
The distribution is plotted as $t_{\rm u} \: ({\rm d} N / {\rm d} t_{\rm u})$
and its downstream equivalent as a function of the dimensionless variable
\[
	\tau \equiv \left\{ \begin{array}{ll}
	{\displaystyle \Gs^{2} {\cal D}_{\theta} t_{\rm u}} & 
	\mbox{upstream} \\
	& \\
	& \\
	\overline{{\cal D}}_{\theta} \overline{t}_{\rm d} &
	\mbox{downstream} \\
	\end{array} \right. \; .
\]
According to our discussion in Section 5 both distributions should peak near
the average value of the residence time which should lie in the range 
$\tau = {\cal O}(1)$.

\begin{figure}
\centerline{\psfig{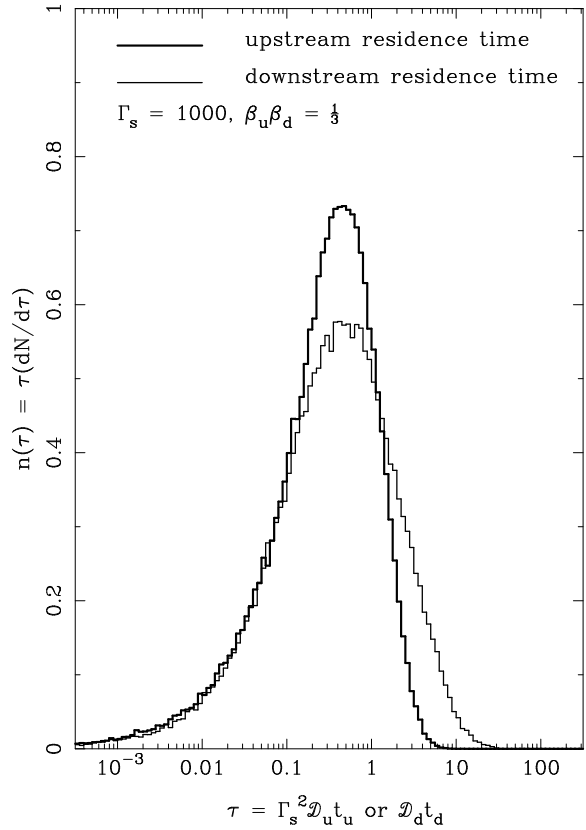}}
\caption{The distribution of up- and downstream residence times for
particles interacting with the shock in the case of isotropic scattering
in both the up- and downstream rest frames. This shows the distribution for
particles with $E > 30 \times E_{\rm inj}$, which ensures that any
influence of the initial conditions has decayed.}
\end{figure}
\nskip
Both distributions peak around $\tau \sim 0.5$, which is the value one
expects if particles have to diffuse through an angle $\Delta \theta \sim
1/\Gs$ upstream, and an angle $\Delta \overline{\theta} \sim 1$ downstream
between shock encounters.

\subsection{Case of regular deflection upstream}

We now turn to the case where upstream particles are deflected by a uniform
regular field $B_{\perp}$.
Table 3 gives the results of our simulations.

\begin{table}
\caption{The spectral slope $s$, the mean ratio of initial and final energy 
$<E_{\rm f}/E_{\rm i}>$ in a crossing cycle upstream $\longrightarrow$ downstream
$\longrightarrow$ upstream,
and the return probability ${\cal P}_{\rm ret}$ for the case of pure deflection
upstream and strong scattering downstream  of the shock.}
\label{scatdefl}
\begin{tabular}{@{}lccc} \hline 
& & & \\
$\Gs$ & $s$ &  $<E_{\rm f}/E_{\rm i}>$ &  ${\cal P}_{\rm ret}$ \\
& & & \\
10 	& 2.28 $\pm$ 0.01 		& 1.67 $\pm$ 0.01 		& 0.516 $\pm$ 0.028 \\
10$^2$	& 2.30 $\pm$ 0.01		& 1.65 $\pm$ 0.01		& 0.519 $\pm$ 0.005 \\
10$^3$	& 2.31 $\pm$ 0.01		& 1.61 $\pm$ 0.01		& 0.533 $\pm$ 0.005 \\
\hline
\end{tabular}
\end{table}

\noindent
The energy gain per crossing is less than in the case of pure scattering,
$\left< \Ef / \Ei \right> \approx 1.64$, but the return probability is higher,
$\Pret \approx 0.523$. As a result, the spectral index of the energy distribution of
accelerated particles is only slighly steeper than in the case of pure scattering:
$s \approx 2.3$. The results are internally consistent: from the average values of
Table 2 we find
\[
	s = 1 + \frac{\ln(1/{\cal P}_{\rm ret})}{\ln \left< \Ef/\Ei \right>}
		= 2.31 \pm 0.02	
\] 

The differences with the case of pure scattering can be explained by looking at
the angular distributions. Figure 8 gives the downstream 
angular distribution for pure scattering up- and downstream, and for the case of 
deflection upstream and scattering downstream.

\begin{figure}
\centerline{\psfig{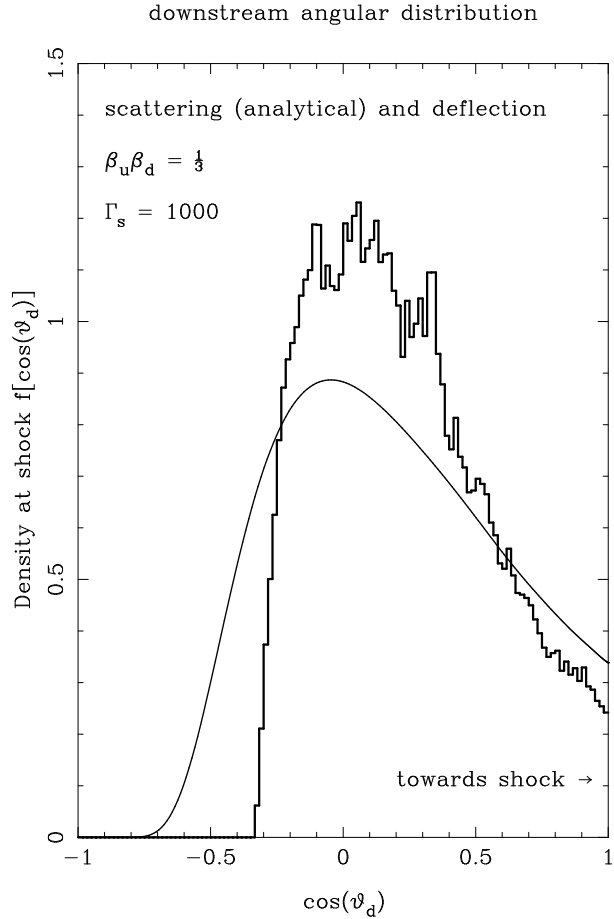}}
\caption{The angular distribution of particles in the downstream rest frame
as a function of $\cos{\theta_{\rm d}} = \overline{\mu}$.
The smooth curve is the distribution of the pure scattering case derived from
the results of Kirk et al. (2000), while the histogram is 
the simulated distribution in the case of regular deflection upstream and 
pure scattering downstream.
The edge of the loss cone is located at $\overline{\mu} = \third$.
In the case of regular deflection upstream, the
distribution extends only to $\overline{\mu} \ge - \third$, 
in agreement with the calculations presented in Section 4, 
Eqn. (\ref{anglimits2}). This curve is for particles with an energy
$E > 30 \times E_{\rm inj}$, where the initial conditions no longer
influence the distribution.
}
\end{figure}

The angular distribution in the case of upstream deflection by a regular field
is narrower, and cuts off at $\mu = - \third$ (see Eqn. \ref{anglimits2}).
This has two effects:
\begin{enumerate}
\item	In the case of regular deflection, the downstream turning angle 
	$\Delta \overline{\theta} = |\bthd - \bthu|$ 
	between shock encounters is smaller, leading
	to a smaller value of $\left< \Ef / \Ei \right>$ in 
	an upstream-downstream-upstream crossing cycle;

\item	Since in this case there is (on average) a smaller fraction of particles 
	with a large value of $\bthd$, the chance $\Pret$ of re-crossing the shock 
	into the upstream medium is larger, and the escape probability $\Pesc$ 
	correspondingly smaller.
\end{enumerate}
This explains the difference of the values in Table 2 and Table 3.
If one plots the energy gain per crossing in a upstream-downstream-upstream 
cycle for both cases (figure 9) the difference is obvious: 
the distribution extends to higher values of $\Ef / \Ei $ 
in the case of pure scattering.

\begin{figure}
\centerline{\psfig{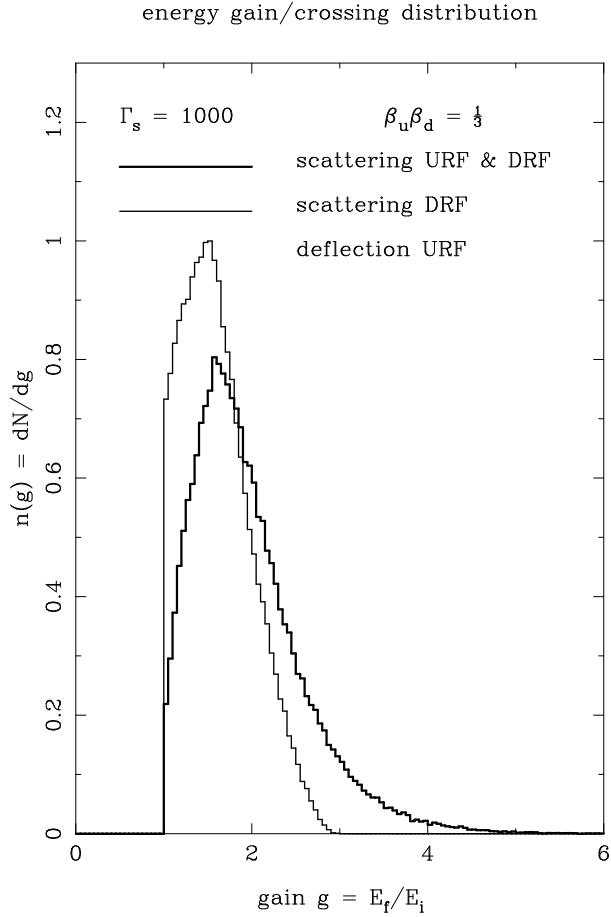}}
\caption{The distribution of the energy gain, $g \equiv \Ef / \Ei$, in an upstream
$\Longrightarrow$ downstream $\Longrightarrow$ upstream crossing cycle.
The curves shown are for particles with $E > 30 \times E_{\rm inj}$ where
the initial conditions no longer influence the dustribution.
The thick line corresponds to the case of pure isotropic scattering on
both sides of the shock, while the thin line 
corresponds to the case where particles are deflected by a regular upstream field,
and are isotropically scattered downstream.
}
\end{figure}

Figure 10 shows the distribution of the upstream residence time $t_{\rm u}(E)$ and
the downstream residence time $\overline{t}_{\rm d}$ for $\Gs = 1000$ and 
upstream deflection, as a function of the dimensionless variable
\[
	\tau \equiv \left\{ \begin{array}{ll}
	{\displaystyle \frac{ZeB_{\perp} \Gs ct_{\rm u}}{E}} & 
	\mbox{upstream} \\
	& \\
	& \\
	\overline{{\cal D}}_{\theta} \overline{t}_{\rm d} &
	\mbox{downstream} \\
	\end{array} \right. \; .
\]
The upstream residence time cuts off sharply at $\tau = 3$, which corresponds
to 
\[
	t_{\rm u} = \frac{3E}{ZeB_{\perp} c \Gs} = 
	\frac{3}{\Omega_{\perp} \Gs} \; ,
\]
the behaviour found in our analytical calculations of Section 4 (see also 
Figure 2). The upstream residence time distribution peaks around 
\[
	t_{\rm u}  \approx \frac{E}{ZeB_{\perp} c \Gs} \; ,	
\]
while downstream there is a rather broad distribution centered around
$\tau \approx 0.2$, which corresponds to
\[
	\overline{t}_{\rm d} \approx 0.2 \: 
	\left(\overline{{\cal D}}_{\theta} \right)^{-1}
	\; ,	
\]
a smaller value for $\overline{t}_{\rm d}$ than obtained in the case of pure
up- and downstream scattering.
This latter result again shows that the average value of
$\Delta \overline{\theta} = |\bthd - \bthu|$ is smaller in the case of regular 
deflection upstream. 

\begin{figure}
\centerline{\psfig{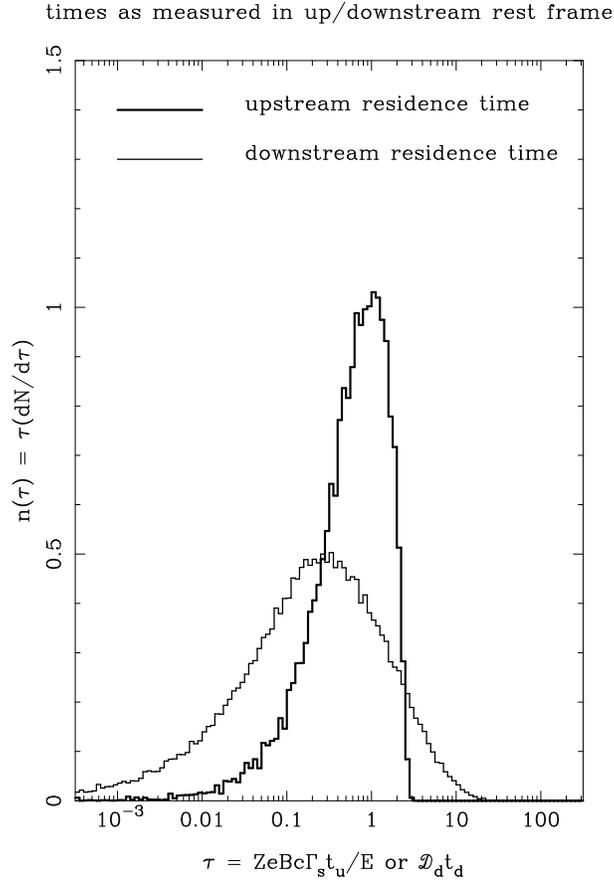}}
\caption{The distribution of up- and downstream residence times for
particles interacting with the shock through deflection by a regular
magnetic field upstream, and isotropic scattering downstream. }
\end{figure}

\subsection{Initial shock encounter}

We have also simulated the initial shock encounter in order to calculate
the energy gain at this first encounter, as well as the return probability
$P_{\rm inj}$ in this case. The latter quantity is essentially the fraction 
of particles picked up for further acceleration, and can therefore be 
considered as an injection probability for true (Fermi-type)
relativistic shock acceleration.  Note that our simulations assume that
the shock itself is infinitely thin which, for these particles, may not
be a very good approximation.  We also assume that the particles are
scattered isotropically in the downstream flow. We have considered two cases:
\nskip
\begin{itemize}
\item	`Cold' upstream particles with $\Ei \approx mc^{2}$ and
	$|\bm{v}| \ll c$;
	
\item	Relativistic particles with $\gamma \gg 1$ distributed isotropically
	in momentum.
\end{itemize}
\nskip
The results in both cases are virtually indistinguishable within the accuracy
allowed by the simulations. 	We find that
\[
	\left< \Ef / \Ei \right> \approx 0.9 \: \Gs^{2} \; \; \; , \; \; \; 
	{\cal P}_{\rm inj} \approx 0.12 \; .
\]
Roughly 10\% of the particles crossing the shock for the first time return	
upstream. A similar result has been found by Bednarz (2000) using a somewhat
different simulation method.

This result is of particular importance for models of Gamma Ray Burst
afterglows. There it is often assumed that {\em all} particles crossing the shock
are accelerated to a power-law distribution. Our results show that this
assumption is probably too optimistic {\em unless} the thermalization of
the bulk of the matter in the shock front also leads to a power-law distribution
in energy for the downstream particles.

\section{Conclusions}

In this paper we have considered the process of shock acceleration
at ultra-relativistic shocks with a shock speed $V_{\rm s} = \bs c$ such that
the corresponding Lorentz factor satisfies $\Gs \gg 1$.

We have shown using both analytical methods and simulations that true 
Fermi-type shock
acceleration, where particles cycle across the shock repeatedly, is subject to
the effects of relativistic beaming. Upstream  particles are confined to an angle
$\theta \lesim 2/\Gs$ with respect to the shock normal.
This limits the energy gain per cycle so that the average
ratio of final and initial
energies in a cycle is $\Ef/\Ei \sim 1.6 - 2.0$, with about 50-60\%
of the particles escaping downstream each cycle.

Like its non-relativistic counterpart, the mechanism is suprisingly robust
in the sense that the energy distribution of the accelerated particles depends
only weakly on the precise mechanism that confines particles near the shock.
Downstream confinement always requires strong scattering in the momentum
direction in order to allow particles to return to the shock. 
Upstream confinement can be achieved in two ways: by isotropic scattering of 
the momentum direction, or by deflection 
by a (quasi-)regular magnetic field. We find that in the first
case the slope of the distribution equals $s \approx 2.23$, while in the latter
case is is somewhat steeper, $s \approx 2.30$. Our results in the case of
up- and downstream isotropic scattering agree with the analytical results of
Kirk et al. (2000), and, in the limit $\Gs \gg 1$, 
with the simulations of Bednarz \& Ostrowski (1998)
who employ a somewhat different simulation method.

The largest particle energies can be achieved in the regime where particles are
deflected rather than scattered upstream. This deflection can be due to
a homogeneous upstream field or due to the sloshing motion in low-frequency 
MHD waves. The maximum energy attainable for a particle with charge $Ze$
at a spherical shock of size $\Rs$ in the absence of losses is
\[
	E_{\rm max} \approx ZeB \: \Gs \Rs \; ,
\] 
where $B$ is the relevant upstream field strength, which is the 
component $B_{\perp}$  of the field perpendicular to the shock normal for a 
homogeneous upstream field, or the magnetic amplitude $\delta B$
of the waves in the case that
the upstream field is closely aligned with the shock normal so that
$B_{\perp} \ll B_{\parallel}/\Gs$, provided this amplitude is sufficiently large:
$\delta B \gesim B_{\parallel}/\Gs$.

The momentum direction upstream will be able to diffuse between shock crossings
if the upstream field is random on a scale $\lc < \Rs$ for particles with an 
energy $E_{\ast} > ZeB \: \Gs \lc$.  
In this case the 
maximum attainable energy, again in the absence of losses, is
\[
	E_{\rm max} \approx ZeB \: \Gs \: \sqrt{\Rs \lc} \; ,
\]
with $B$ the typical amplitude of the random field.

The above two limits apply mostly to hadrons which suffer little or
no radiation losses. For electrons (or positrons) synchrotron losses
can limit the maximum energy to an energy of order
\[
	E_{\rm max} \approx m_{\rm e} c^{2} \: 
	\left( \frac{4 \pi e \Gs}
	{\sigma_{\rm T} \xi_{\rm B} \: B_{\perp}} \right)^{1/2} \; ,  
\]
where $\xi_{\rm B}$ is the downstream field amplification factor defined in
Eqn. (\ref{amppar}). All these energies are measured in the URF.

The critical frequency $\nu_{\rm s}$ of the synchrotron radiation generated
by  electrons with Lorentz factor $\gamma_{\rm e} = E/m_{\rm e} c^{2}$
is (Rybicki \& Lightman, 1979, Ch. 6)
\[
	\nu_{\rm s} \approx \frac{3}{4 \pi} \:
	\left( \frac{eB_{\perp}}{m_{\rm e}c} \right) \: \gamma_{\rm e}^{2}  \; .
\]
If synchrotron losses determine $E_{\rm max}$,  
particles with the maximum energy radiate photons with energy
\[
	h \nu_{\rm max} \approx \frac{9 \Gs}{8 \pi} \: 
	\left( \frac{m_{\rm e}c^{2}}{\alpha} \right) \approx 
	25 \: \Gs \; {\rm MeV}\; , 
\]
in the URF,
with $\alpha = e^{2}/\hbar c \approx 1/137.04$ the fine-structure constant.
This limit (roughly) applies to synchrotron photons originating in the upstream
flow, and to the Doppler-boosted photons originating in the downstream flow.
Note that the maximum photon energy is independent of the magnetic field
strength in the source.

We also considered the effect of the shock on particles at the first encounter. 
Assuming an infinitely thin shock with isotropic scattering downstream, we find 
that about 10\% of the particles manage to return upstream so that they can 
participate in the shock acceleration process. They return upstream with an 
energy $E \sim 0.9 \Gs^{2} E_{0}$, where $E_{0}$ is their energy before the 
encounter with the shock. This provides a natural injection process for 
shock acceleration at ultra-relativistic shocks.

We note here that electrons will have to be pre-accelerated in order to
be picked up by the shock acceleration process, assuming the thickness of 
the (collisionless) shock is of order 
the gyration radius of the downstream protons, which have an energy
\[
	\overline{E}_{\rm p} \sim \Gs m_{\rm p} c^{2}
\]
in the downstream rest frame.
Cold upstream electrons entering the downstream flow have an energy 
$\overline{E}_{\rm e} \sim \Gs m_{\rm e} c^{2}$, and must be 
pre-accelerated by a different mechanism 
up to proton energies before they will start to see 
the shock as a discontinuity. A possible mechanism which could provide the
necessary pre-acceleration has been 
proposed in the context of relativistic shocks in a pulsar wind by
Hoshino {\em et al.} (1992).	

\section*{Acknowledgements}

This work is a collaboration within the Astroplasmaphysics Network,
supported under the TMR programme of the European Union, 
contract ERBFMRXCT98-0168.
Y.A.G. acknowledges support from the Netherlands Foundation for Research
in Astronomy (ASTRON project 781--76--014).


\begin{thebibliography}{}

\bibitem[]{AchtKr}
Achterberg, A. \& Kr\"ulls, W.M. 1992, A\&A, 265, L13.

\bibitem{AchtEADN}
Achterberg, A., 1993, in:\\ 
{\em Galactic High-Energy Astrophysics/High-Accuracy Timing
and Positional Astronomy}, J. van Paradijs \& H.M. Maitzen (Eds.), 
Springer Verlag, Heidelberg, p. 4.


\bibitem{ABlRe}
Achterberg, A., Blandford, R.D., Reynolds, S.P., 1994, 
Astron. Astrophys. 281, 220, 1994.

\bibitem{AchtBall}
Achterberg, A., Ball, L., 1994, A\&A 284, 687
\bibitem{Achtetal}

Achterberg, A., Gallant, Y.A., Norman,C.A., Melrose, D.B., 1999,
submitted to MNRAS, astro-ph/9907060

\bibitem{AsLe}
Aschenbach, B., Leahy, D.A., 1999, A\&A 341, 602.

\bibitem{ALS} Axford, W.I., Leer, E. \& Skadron, G., 1978, 
{\em Proc. 15 Int. Cosmic Ray Conf.} (Plovdiv) 11, 132.

\bibitem{BaHe}
Ballard, K.R. and Heavens, A.F. 1992, MNRAS 259, 89

\bibitem{Bell} 
Bell, A.R., 1978, MNRAS 182, 147.

\bibitem{BeOs}
Bednarz, J. \& Ostrowski, M., 1998, Phys. Rev. Lett. 80, 3911

\bibitem{Bedn}
Bednarz, J., 2000, MNRAS 315, L37

\bibitem{BhaSi}
Bhattacharjee, P., Sigl, G., 2000, Phys. Rep. 327, 109

\bibitem{Bird}
Bird, D.J. \etal 1994, ApJ, 424, 491

\bibitem{Bla79} 
Blandford, R.D. 1979, Proc. AIP Conf (La Jolla), 56, 335

\bibitem{BlaMcK76}
Blandford R. D., McKee C. F., 1976, Phys.\ Fluids, 19, 1130


\bibitem{BlaOst78}
Blandford R.D., Ostriker J.P., 1978, ApJ, 221, L29.

\bibitem{BlEi}
Blandford, R.D., Eichler, D., 1987, Phys. Rep. 154,

\bibitem{CaRe}
Cavallo, G., Rees, M.J., 1987, MNRAS, 183, 359

\bibitem{Djoretal}
Djorgowski, S.G. \etal 1997, Nature, 387, 876

\bibitem{Drurev}
Drury, L.O'C., 1983, Rep. Prog. Phys. 46, 963.

\bibitem{FaBie}
Farrar, G.R. \& Biermann, P.L. 1998, Phys. Rev Lett. 81, 3579.

\bibitem{Fe}
Fermi, E., 1949, Phys. Rev. 75, 1169.

\bibitem{FKNFT}
Frail, D.A., Kulkarni, S.R., Nicastro, L., Feroci, M., Taylor, G.B., 1997,
Nature, 389, 261

\bibitem{GaAcht}
Gallant, Y.A., Achterberg, A., 1999, MNRAS, 305, L6

\bibitem[]{Gard}
Gardiner, C.W. 1983, {\em Handbook of Stochastic Methods}, 
Springer Verlag, Berlin.

\bibitem{Go}
Goodman, J., 1986, ApJ 308, L47

\bibitem{Grei}
Greisen, K. 1966, Phys. Rev. Lett., 16, 748.

\bibitem{He}
Heavens, A.F. 1984, MNRAS 207, 1P

\bibitem{HeDr}
Heavens, A.F., Drury, L.O'C. 1988, MNRAS 235, 997

\bibitem{Hi}
Hillas, A.M., 1984, Ann. Rev. Astron. Astroph. 22, 425.

\bibitem{Hosetc92}
Hoshino M., Arons J., Gallant Y.A., Langdon A.B., 1992, ApJ 390, 454

\bibitem{Jok}
Jokipii, J.R., 1987, ApJ 313, 842.

\bibitem{JoEl}
Jones, F.C., Ellison, D.C., 1991, Space Sc. Rev. 58, 259.

\bibitem{KRJ}
Kang, H., Ryu. D., Jones, T.W., 1996, ApJ 456, 422.

\bibitem{Kirev}
Kirk, J.G., 1994, in: {\em Plasma Astropysics}, 
Saas-Fee Advanced Course 24, 
A.O. Benz \& T.J.-L. Courvoisier (Eds.), 
Springer Verlag, Heidelberg, p. 225. 

\bibitem{KiSch1}
Kirk, J.G., Schneider, P., 1987a, ApJ 315, 425.

\bibitem{KiSch2}
Kirk, J.G., Schneider, P., 1987b, ApJ 322, 256.


\bibitem{KiDu}
Kirk, J.G., Duffy, P., 1999, J. Phys. Nucl. Part. Phys. 25, R163

\bibitem{KiGu}
Kirk, J.G., Guthmann, A.W., Gallant, Y.A., Achterberg, A, 2000,
ApJ 542, 235

\bibitem{Koy}
Koyama, K., Petre, R., Gotthelf, E.V., Hwang, U., Matsuura, M., 
Ozaki, M., Holt, S.S., 1995, Nature 378, 255.

\bibitem{Kry77}
Krymskii G.F., 1977, Sov.\ Phys.\ Dokl., 22, 327

\bibitem{LaCe}
Lagage, P.O., Cesarsky, C.J. 1983, A\&A, 118, 223

\bibitem{MeRe1}
M\'esz\'aros, P., Rees, M.J., 1992a, ApJ, 397, 570

\bibitem{MeRe2}
M\'esz\'aros, P., Rees, M.J., 1992b, MNRAS, 257, 29P

\bibitem{MeRe3}
M\'esz\'aros, P., Rees, M.J., 1992c, MNRAS, 258, 41P

\bibitem{MeRe4}
M\'esz\'aros, P., Rees, M.J., 1993, ApJ, 405, 278

\bibitem{Metzetal}
Metzger, M.R., Djorgovski, S.G., Kulkarni, S.R., Steidel, C.C.,
Adelberger, K.L., Frail, D.A., Costa, E., Frontera, F., 1997,
Nature 387, 878

\bibitem{MiUs}
Milgrom, M., Usov, V. 1995, ApJ, 449, L37


\bibitem{NoMeAcht}
Norman, C.A., Melrose, D.B., Achterberg, A. 1995, ApJ, 454, 60

\bibitem[]{Oks}
{\O}ksendal, B. 1991, {\em Stochastic Differential Equations}, 
third ed., Springer Verlag, Berlin.

\bibitem{Pa}
Paczy\'nski, B., 1986, ApJ, 308, L43

\bibitem{Peac}
Peacock, J.A., 1981, MNRAS 196, 135. 

\bibitem{PiShNa}
Piran, T., Shemi, A., Narayan, R., 1993, MNRAS, 263, 861

\bibitem{Pi}
Piran, T., 1999, Phys. Rep. 314, 575

\bibitem{RaBie}
Rachen, J.P., Biermann, P.L. 1993, A\&A, 272, 161

\bibitem{ReMe}
Rees, M.J., M\'esz\'aros, P., 1994, ApJ, 430, L93

\bibitem{RyLi}
Rybicki, G.B., Lightman, A.P., 1979, {\em Radiative Processes in\\
Astrophysics}, Cambridge Univ. Press.

\bibitem{SaPi1}
Sari, R., Piran, T., 1995, ApJ, 455, L143

\bibitem{SaPi2}
Sari, R., Piran, T., 1997, ApJ, 485, 270


\bibitem{SaPiNa}
Sari, R., Piran, T., Narayan, R., 1998, ApJ Lett. 497, L17.

\bibitem{SchKi}
Schneider, P., Kirk, J.G., 1989, Astron. Astrophys. 217, 344. 

\bibitem{ShPi}
Shemi, A., Piran, T., 1990, ApJ, 365, L55

\bibitem{SiSchBa}
Sigl, G., Schramm, B.N. \& Bhattacharjee, P. 1994, Astrop. Phys., 2, 401.


\bibitem{Taketal}
Takeda, M. \etal 1998, Phys. Rev. Lett. 81, 1163


\bibitem{vParetal}
Van Paradijs, J. \etal 1997, Nature, 386, 686

\bibitem{Vie}
Vietri, M. 1995, ApJ, 453, 883

\bibitem{Wa1}
Waxman, E. 1995, Phys. Rev. Lett., 75, 386

\bibitem{Webb}
Webb, G.M. 1985, ApJ, 296, 319

\bibitem{We74}
Wentzel, D.G. 1974, Ann. Rev. Astron. Astrophys. 12, 71

\bibitem{Yosetal}
Yoshida, S. \etal 1995, Astrop. Phys., 3, 105

\bibitem{ZaKu}
Zatsepin, G.T. \& Kuz'min, V.A. 1966, JETP Lett., 4, 78.

\end{thebibliography}
\end{document}